\documentclass[useAMS]{mn2e}

\usepackage{graphicx}
\usepackage{amsmath}
\usepackage{amsfonts}
\usepackage{amssymb}
\usepackage{rotating}

\def\aj{AJ}%
\def\araa{ARA\&A}%
\def\apj{ApJ}%
\def\apjl{ApJ}%
\def\apjs{ApJS}%
\def\aap{A\&A}%
\def\mnras{MNRAS}%

\title{Smoothed  Particle  Hydrodynamics  simulations of  white  dwarf
       collisions and close encounters}

\author[P. Lor\'en--Aguilar et al.]{P. Lor\'en--Aguilar$^{1,2}$,
                                    J. Isern$^{3,2}$ and
                                    E. Garc\'\i a--Berro$^{1,2}$\\
       $^1$Departament de F\'\i sica Aplicada, 
           Universitat Polit\`ecnica de Catalunya,
           c/Esteve Terrades 5, 
           08860 Castelldefels, Spain\\
       $^2$Institute for Space  Studies of Catalonia,
           c/Gran Capit\`a 2--4, Edif. Nexus 104,   
           08034  Barcelona,  Spain\\
       $^3$Institut de Ci\`encies de l'Espai, CSIC,  
           Campus UAB, Facultat de Ci\`encies, Torre C-5, 
           08193 Bellaterra, Spain}

\begin{document}

\date{\today}

\maketitle

\begin{abstract}
The collision of  two white dwarfs is a quite  frequent event in dense
stellar systems,  like globular clusters and galactic  nuclei. In this
paper we  present the  results of  a set of  simulations of  the close
encounters and  collisions of two  white dwarfs. We use  an up-to-date
smoothed particle  hydrodynamics code that  incorporates very detailed
input physics  and an improved treatment of  the artificial viscosity.
Our  simulations have  been done  using  a large  number of  particles
($\sim  4\times10^5$) and  covering  a wide  range  of velocities  and
initial distances of the colliding white dwarfs.  We discuss in detail
when  the initial eccentric  binary white  dwarf survives  the closest
approach,  when a  lateral collision  in which  several  mass transfer
episodes occur is  the outcome of the newly  formed binary system, and
which range of input parameters  leads to a direct collision, in which
only  one   mass  transfer  episode  occurs.   We   also  discuss  the
characteristics of the final  configuration and we assess the possible
observational  signatures  of  the  merger,  such  as  the  associated
gravitational waveforms  and the fallback luminosities.   We find that
the overall  evolution of the  system and the main  characteristics of
the final object agree with  those found in previous studies.  We also
find  that the  fallback luminosities  are close  to  $10^{48}$ erg/s.
Finally,  we find  as well  that  in the  case of  lateral and  direct
collisions   the   gravitational   waveforms  are   characterized   by
large-amplitude peaks  which are followed by a  ring-down phase, while
in  the case  in which  the binary  white dwarf  survives  the closest
approach,  the  gravitational pattern  shows  a distinctive  behavior,
typical of eccentric systems.
\end{abstract}



\section{Introduction}

In recent  years, the study  of stellar collisions has  attracted much
interest from  the astronomical community  working on the  dynamics of
dense  stellar  systems,  like  the  cores of  globular  clusters  and
galactic nuclei (Shara  2002). One of the reasons for  this is that in
these systems,  stellar collisions are  rather frequent (Hills  \& Day
1976).  In fact, it has been predicted that up to 10\% of the stars in
the core  of typical globular  clusters have undergone a  collision at
some point during the lifetime of the cluster (Davies 2002).

The most  probable collisions are those  in which at least  one of the
colliding stars has the largest possible cross section --- a red giant
--- and those in which at least one of the stars is most common (Shara
\&  Regev 1986).   This later  type of  collisions  obviously includes
those in  which a  main sequence star  is involved.   However, because
white dwarfs  are the most common  end point of  stellar evolution and
because  both globular clusters  and galactic  nuclei are  rather old,
these   stellar  systems   contain  many   collapsed   and  degenerate
objects.  Therefore, we  expect that  collisions in  which one  of the
colliding stars is a white dwarf should be rather common.

Collisions  between  two  main  sequence  stars  are  supposed  to  be
responsible for the observed population of blue stragglers in globular
clusters (Sills  \& Bailyn  1999; Sills et  al.  2005).   Although the
collision  between  a main  sequence  star and  a  red  giant is  more
probable than  that of two main  sequence stars because  of the larger
geometrical cross section of the  red giant star, they probably do not
produce an  interesting astrophysical object.  The reason  for this is
the low density of the envelopes  of red giants. In most cases, during
the encounter the red giant is deprived of part of its envelope and is
able to recover its appearance (Freitag \& Benz 2005).  The collisions
of a white dwarf  and a red giant or a main  sequence star are also of
large interest,  since they  may be responsible  for the  formation of
some  interesting astrophysical  objects.  Unfortunately,  due  to the
very   different  dynamical   scales   involved,  the   hydrodynamical
simulation of  these events is difficult and  realistic simulations of
their outcome are  still lacking. However, in the case  in which a red
giant and a  white dwarf collide it is thought  that the most probable
outcome  is the  ejection of  the envelope  of the  red giant  and the
formation of a double white dwarf binary system (Tuchman 1985), whilst
in the case in which a main sequence star and a white dwarf collide it
has been shown (Shara \& Regev 1986) that only a small fraction of the
disrupted main  sequence star remains  bound to the white  dwarf. More
recent  simulations (Ruffert  1992) predict  the formation  of  a disk
around  the  white  dwarf.   However,  we  emphasize  that  all  these
simulations used rudimentary input  physics and, thus, the outcomes of
these simulations are dubious.

The collision of two white  dwarfs deserves study for various reasons.
In particular, the collision of two white dwarfs can produce a Type Ia
supernova. Although the most standard  scenario for a Type Ia outburst
--- the  so-called  single-degenerate scenario  ---  involves a  white
dwarf accreting from a non-degenerate companion, the double-degenerate
scenario (Webbink 1984; Iben \& Tutukov 1984), in which the merging of
two  carbon-oxygen white  dwarfs with  a  total mass  larger than  the
Chandrasekhar limit occurs, has been one of the most favored scenarios
leading to  Type Ia supernovae.  In  fact, it has  been predicted that
the white  dwarf merger  rate leading to  super-Chandrasekhar remnants
will  be  increased  by   an  order  of  magnitude  through  dynamical
interactions  (Shara \&  Hurley  2002). Therefore,  collisions of  two
white  dwarfs of  sufficiently large  masses could  explain supernovae
occurring in the  nuclei of galaxies.  Moreover, it  has been recently
suggested that such  a process would lead to both  a Type Ia supernova
explosion  and  to the  formation  of  a  magnetar (King,  Pringle  \&
Wickramasinghe   2001).   This   scenario  would   explain   the  main
characteristics  of  soft  gamma-ray  repeaters  and  anomalous  X-ray
pulsars  like 1E2259+586.   Also, dynamical  interactions  in globular
clusters can  form double  white dwarfs with  non-zero eccentricities,
which would be powerful sources of gravitational radiation (Willems et
al. 2007). Moreover, the initial  stages of the coalescence of a white
dwarf binary system  could be one of the  most interesting sources for
the detection of gravitational  waves using space-borne detectors like
LISA ({\tt http://lisa.jpl.nasa.gov}).  Thus, a close encounter of two
white dwarfs would be a potentially observable source of gravitational
waves and,  hence, characterizing the gravitational  waveforms is also
of interest. Finally, the  temperatures achieved in a direct collision
are substantially high  and, consequently, we expect that  some of the
nuclearly processed material could  be ejected, leading to a pollution
of the  environment where  it occurs, either  a globular cluster  or a
galactic nucleus.

Despite of its  potential interest, there are very  few simulations of
the collision of  two white dwarfs, the only  exception being those of
Benz et  al. (1989), Rosswog et  al. (2009) and Raskin  et al. (2009).
All  three sets  of simulations  used Smoothed  Particle Hydrodynamics
(SPH) to  model the collisions.   However, the simulations of  Benz et
al.  (1989) were done using  a small number of particles whereas those
of Rosswog  et al.  (2009)  and Raskin et  al.  (2009) have  studied a
limited range of impact parameters. To be specific, the simulations of
Benz  et al.   (1989) used  $5\times 10^3$  particles, while  those of
Rosswog et  al. (2009)  and Raskin et  al. (2009)  used, respectively,
$2.5\times  10^6$ and $8\times  10^5$ SPH  particles. The  very recent
simulations of  Rosswog et al.  (2009)  and Raskin et  al. (2009) were
aimed  to  produce a  thermonuclear  explosion  and,  thus, they  only
studied  direct  collisions, while  the  simulations  of  Benz et  al.
(1989) covered a broader range of initial conditions. Additionally, in
the early work of Benz et al.  (1989) the classical expression for the
artificial viscosity (Monaghan \& Gingold 1983) was used, while in the
very  recent calculations  of Rosswog  et  al.  (2009)  and Raskin  et
al. (2009) more elaborated  prescriptions for the artificial viscosity
were employed.

In  sharp  contrast,  the  coalescence  of  binary  white  dwarfs  was
extensively  studied in  the  past and  also  has been  the object  of
several  recent  studies.   For  instance,  the  pioneering  works  of
Mochkovitch \& Livio  (1989, 1990) used an approximate  method --- the
so-called  Self-Consistent-Field method (Clement  1974) ---  while the
full  SPH  simulations of  Benz,  Thielemann  \&  Hills (1989),  Benz,
Cameron \&  Bowers (1989), Benz,  Hills \& Thielemann (1989),  Benz et
al.   (1990),  Rasio \&  Shapiro  (1995)  and  Segretain, Chabrier  \&
Mochkovitch (1997)  studied the problem using  reduced resolutions and
the  classical expression  for the  artificial viscosity  (Monaghan \&
Gingold 1983).  Later, Guerrero et  al.  (2004) opened the way to more
realistic simulations, using an  increased number of SPH particles and
an improved prescription for the artificial viscosity.  More recently,
the  simulations of  Yoon et  al.  (2007) and  of Lor\'en--Aguilar  et
al. (2005, 2009)  were carried out using modern  prescriptions for the
artificial viscosity and even larger numbers of particles. All in all,
it is noticeable the lack of SPH simulations of white dwarf collisions
and  close encounters  when compared  to the  available  literature on
white dwarf mergers.

In  the present  paper  we study  the  collision of  two white  dwarfs
employing   an  enhanced  spatial   resolution  ($4\times   10^5$  SPH
particles) and  an improved formulation for  the artificial viscosity.
We pay  special attention to  discern the range of  initial conditions
that produce the  tidal disruption of the less  massive white dwarf or
those  for which  the initial  eccentric binary  survives  the closest
approach.  The  number of  particles used in  our simulations  is much
larger than those  used in the simulations of Benz  et al.  (1989) and
in line with  those used in modern simulations  (Rosswog et al.  2009;
Raskin  et al.  2009).   However, our  calculations encompass  a broad
range of initial conditions of the colliding white dwarfs, in contrast
to most modern simulations, in which  only a few cases were studied in
detail.  The paper  is organized as follows.  In \S  2 we describe our
input physics and the  method of calculation, paying special attention
to describe with some detail our  SPH code.  It follows \S 3, which is
devoted  to discuss  the  initial conditions  adopted  in the  present
study,  while in  \S 4  we describe  the results  of  our simulations.
Finally  in  \S  4  we  summarize  our  main  findings  and  draw  our
conclusions.


\section{Input physics and method of calculation}

We follow  the hydrodynamic evolution of the  interacting white dwarfs
using  a Lagrangian  particle numerical  code, the  so-called Smoothed
Particle Hydrodynamics.  This method was first proposed by Lucy (1977)
and, independently, by Gingold \&  Monaghan (1977).  The fact that the
method  is totally Lagrangian  and does  not require  a grid  makes it
especially  suitable for  studying an  intrinsically three-dimensional
problem like the collision of  two white dwarfs.  We will not describe
in detail the  most basic equations of our  numerical code, since this
is a  well-known technique.  Instead,  the reader is referred  to Benz
(1990) where the  basic numerical scheme for  solving the hydrodynamic
equations  can be  found, whereas  a general  introduction to  the SPH
method  can be  found  in  the excellent  review  of Monaghan  (2005).
However, and  for the  sake of completeness,  we briefly  describe the
most relevant equations of our numerical code.

We use the standard polynomic  kernel of Monaghan \& Lattanzio (1985).
The gravitational forces are evaluated  using an octree (Barnes \& Hut
1986). Also,  gravitational forces between SPH  particles are softened
using  the procedures  described  in Monaghan  \&  Gingold (1977)  and
Hernquist \&  Katz (1989).  Our SPH  code uses a  prescription for the
artificial  viscosity   based  in  Riemann-solvers   (Monaghan  1997).
Additionally, to  suppress artificial  viscosity forces in  pure shear
flows, we  also use the viscosity  switch of Balsara  (1995).  In this
way the  dissipative terms  are largely reduced  in most parts  of the
fluid and are  only used where they are really  necessary to resolve a
shock, if  present.  Within this  approach, the SPH equations  for the
momentum and energy conservation read, respectively, as

\begin{eqnarray}
\frac{d\vec{v}_i}{dt} =&-& \sum_j m_j \left( \frac{P_i}{\rho_i^2} + 
\frac{P_j}{\rho_j^2} - \alpha \overline{f}_{ij}\frac{v^{\rm sig}_{ij}}
{\overline{\rho}_{ij}}\vec{v}_{ij}\cdot \hat{e}_{ij} \right) \nonumber\\
&\,&\vec{r}_{ij} \overline{F}_{ij}\\
\frac{du_i}{dt} =&\,& \frac{P_i}{\rho_i^2} \sum_j m_j \vec{v}_{ij} \cdot
\vec{r}_{ij} \overline{F}_{ij}\nonumber \\ 
&-& \frac{1}{2}\sum_j m_j \alpha \overline{f}_{ij}\frac{v^{\rm sig}_{ij}}
{\overline{\rho}_{ij}}(\vec{v}_{ij} \cdot \hat{e}_{ij})^2 
|\vec{r}_{ij}| \overline{F}_{ij}
\label{sph}
\end{eqnarray}

\noindent where $\overline{f}_{ij} = (f_i + f_j)/2$ and
\begin{equation}
f_i = \frac{|\nabla \cdot \vec{v}|_i}{|\nabla \cdot \vec{v}|_i + 
|\nabla \times \vec{v}|_i + 10^{-4}{c_i}/{h_i}},
\end{equation}
In  these  expressions  $\vec{r}_{ij}  = \vec{r}_{i}  -  \vec{r}_{j}$,
$\vec{v}_{ij}   =   \vec{v}_{i}   -  \vec{v}_{j}$,   $\hat{e}_{ij}   =
\vec{r}_{ij}/   |\vec{r}_{ij}|$,   $\overline{\rho}_{ij}=  (\rho_i   +
\rho_j)/2$, $\overline{F_ {ij}}= (F_i+F_j)  /2$, and $F$ is a function
that only  depends on  $|\vec{r}|$ and on  the smoothing  kernel $h_i$
used  to express  the gradient  of the  kernel $\vec{\nabla}  W_{ij} =
\overline{F}_{ij}\vec{r}_{ij}$.   The rest of  the symbols  have their
usual meaning.   The signal velocity  is taken as $v^{\rm  sig}_{ij} =
c_i+ c_j  -4\vec{v}_{ij}\cdot \hat{e}_{ij}$, where $c_i$  is the sound
speed of  particle $i$.   Note that in  the expression for  the signal
velocity  we have arbitrarily  fixed the  coefficient of  the relative
velocity, $\vec{v}_{ij}$, to the value recommended by Monaghan (2005).
We find that $\alpha=0.5$ yields good results.

We  have found  that  it is  sometimes  advisable to  use a  different
formulation of  the equation of energy  conservation. Accordingly, for
each time step  we compute the variation of  the internal energy using
Eq.~(\ref{sph})   and  simultaneously   calculate  the   variation  of
the temperature using

\begin{eqnarray}
\frac{dT_{i}}{dt}=&-&\sum_{j}
\frac{m_j}{(C_v)_j}\frac{T_j}{\rho_i\rho_j}
\left[\left(\frac{\partial P}{\partial T}\right)_\rho\right]_j
\vec{v}_{ij}\cdot \vec{r}_{ij}\overline{F}_{ij}\nonumber\\
&-& \frac{1}{2}\sum_j \frac{m_j}{(C_v)_j} \alpha
\overline{f}_{ij}\frac{v^{\rm sig}_{ij}}
{\overline{\rho}_{ij}}(\vec{v}_{ij} \cdot \hat{e}_{ij})^2
|\vec{r}_{ij}| \overline{F}_{ij}
\label{temp}
\end{eqnarray}

\noindent  where $C_{v}=(\partial U/\partial  T)_{V}$ is  the specific
heat  capacity per  unit  mass ---  see  Timmes \&  Arnett (1999)  and
Segretain et al. (1994) and  references therein for more details about
the implementation of the equation of state.  For regions in which the
temperatures are  lower than  $6 \times 10^8$  K or the  densities are
lower  than  $6 \times  10^3$  g/cm$^3$,  Eq.~(\ref{sph}) is  adopted,
whereas Eq.~(\ref{temp}) is  used in the rest of  the fluid.  We adopt
this  procedure because  the internal  energy of  degenerate electrons
depends very  weakly on  the temperature.  Thus,  in the  region where
degeneracy  is  large small  variations  of  the  internal energy  can
produce  large  fluctuations  of  the  temperature.  The  use  of  Eq.
(\ref{temp}) avoids numerical artifacts  and allows to use longer time
steps.  Using this prescription we find that energy is best conserved.
Specifically, we find that, depending  on the run, energy is conserved
to  accuracies  ranging from  0.1\%  to  3.2\%.   Angular momentum  is
conserved to an accuracy of 0.1\% in the worst of the cases.

\begin{table*}
\begin{center}
\begin{tabular}{ccccccccccccc}
\hline
\hline
Run &
$y_{\rm ini}$ &
$v_{\rm ini}$ &
Outcome &
$E$ &
$L$ &
$r_{\max}$ &
$r_{\min}$ &
$\varepsilon$ &
$\beta$\\
&
($R_{\sun}$) &
(km/s) &
&
($10^{48}$ erg) &
($10^{50}$ erg/s) &
(0.1 $R_{\sun}$) &
(0.1 $R_{\sun}$)&
\\
\hline
   1  & 0.8 & 100 & O  & $-2.13$ & 7.49 & 8.28 & 0.50 & 0.886 & 0.40 \\
   2  & 0.5 & 150 & O  & $-3.17$ & 7.02 & 5.46 & 0.45 & 0.848 & 0.44 \\
   3  & 0.5 &  50 & DC & $-3.44$ & 2.33 & 5.39 & 0.05 & 0.983 & 4.00 \\
   4  & 0.3 & 225 & O  & $-4.49$ & 6.31 & 3.80 & 0.37 & 0.823 & 0.54 \\
   5  & 0.3 & 200 & LC & $-4.64$ & 5.62 & 3.75 & 0.29 & 0.858 & 0.69 \\
   6  & 0.3 & 175 & LC & $-4.77$ & 4.91 & 3.71 & 0.21 & 0.891 & 0.95 \\
   7  & 0.3 & 150 & LC & $-4.88$ & 4.21 & 3.68 & 0.16 & 0.919 & 1.25 \\
   8  & 0.3 & 125 & LC & $-4.98$ & 3.51 & 3.66 & 0.11 & 0.943 & 1.82 \\
   9  & 0.3 & 100 & DC & $-5.05$ & 2.81 & 3.64 & 0.07 & 0.964 & 2.86 \\
  10  & 0.1 & 200 & DC & $-7.82$ & 1.87 & 2.36 & 0.03 & 0.975 & 6.67 \\
  11  & 0.1 & 150 & DC & $-8.06$ & 1.40 & 2.31 & 0.02 & 0.986 & 10.0 \\
  12  & 0.1 & 120 & DC & $-8.16$ & 1.17 & 2.28 & 0.01 & 0.990 & 20.0 \\
\hline
\hline
\end{tabular}
\end{center}
\caption{Summary  of  the  kinematical  properties of  the  $0.6+0.8\,
         M_{\sun}$  simulations.  Note  that  the radius  of the  less
         massive white dwarf is  $R_2\sim 0.01\, R_{\sun}$. Energy and
         angular momentum  have been calculated in the  center of mass
         frame.  The  results presented  here  correspond  to the  SPH
         calculations. }
\label{outcome}
\end{table*}

The algorithm used to determine  the smoothing length of each particle
is that  of Hernquist \&  Katz (1989). That  is, we determine  the new
smoothing  length taking into  account the  previous one  and imposing
that the new  one should be such that the  average number of neighbour
particles  should remain  constant. In  our calculations  we  adopt 32
neighbour   particles.   For   the  integration   method,  we   use  a
predictor-corrector numerical scheme  with variable time steps (Serna,
Alimi \&  Chieze 1996),  which turns out  to be quite  accurate.  Each
particle  is followed  with  individual time  steps.   Time steps  are
determined  comparing   the  local  sound  velocity   with  the  local
acceleration and  imposing that  none of the  SPH particles  travels a
distance  larger  than its  corresponding  smoothing  length. We  also
impose that the temperature or the energy do not vary in one time step
by more than 5\%.  

The equation of state adopted for  the white dwarf is the sum of three
components.  The  ions are  treated as  an ideal gas  but we  take the
Coulomb corrections  into account (Segretain  et al.  1994).   We have
also  incorporated the  pressure of  photons,  which turns  out to  be
important when the temperature is  high and the density is small, just
when nuclear  reactions become  relevant.  Finally the  most important
contribution is the pressure of degenerate electrons, which is treated
by integrating the Fermi-Dirac integrals.  The nuclear network adopted
here incorporates 14 nuclei: He, C, O,  Ne, Mg, Si, S, Ar, Ca, Ti, Cr,
Fe, Ni,  and Zn.   The reactions considered  are captures  of $\alpha$
particles, and  the associated  back reactions, the  fussion of  two C
nuclei, and  the reaction between C  and O nuclei.  All  the rates are
taken  from  Rauscher \&  Thielemann  (2000).   The screening  factors
adopted in  this work are  those of Itoh  et al. (1979).   The nuclear
energy release  is computed  independently of the  dynamical evolution
with much shorter time steps, assuming that the dynamical variables do
not  change much during  these time  steps.  Finally,  neutrino losses
have  also been  included  according  to the  formulation  of Itoh  et
al. (1996)  for the pair,  photo, plasma, and  bremsstrahlung neutrino
processes.

In order to  explore a wide range of input  parameters we have relaxed
two initial  carbon-oxygen white  dwarf models with  masses $M_1=0.8\,
M_{\sun}$ and $M_2=0.6\, M_{\sun}$, respectively. Carbon and oxygen in
these models have equal mass abundances and are uniformly distributed.
The  temperature  of our  isothermal  white  dwarf  initial models  is
$T=10^7$ K,  a rather typical  value.  To achieve  equilibrium initial
configurations, each individual model  star was relaxed separately, so
the  two interacting  white dwarfs  are spherically  symmetric  at the
beginning  of  our  simulations,  as  was the  case  in  all  previous
simulations  of  this  kind.   Finally,  we emphasize  that  to  avoid
numerical   artifacts,   we  only   use   equal-mass  SPH   particles.
Consequently, the $0.8\, M_{\sun}$ white dwarf was relaxed using $\sim
2.6 \times  10^{5}$ SPH particles, whereas the  $0.6 \,M_{\sun}$ white
dwarf  needed $\sim  2.0 \times  10^{5}$ SPH  particles.   However, we
would  like to note  that we  also performed  some additional  runs in
which  the number  of particles  was a  factor of  10 smaller,  and we
obtained essentially the same results.  Neverteless, these simulations
are not presented here. The central densities of the white dwarfs are,
respectively, $\rho_1\simeq 1.0\times 10^7$ g/cm$^3$ and $\rho_2\simeq
3.6 \times  10^6$ g/cm$^3$, while their respective  moments of inertia
(which  are important for  discussing of  our results)  are $I_1\simeq
2.50 \times  10^{50}$ g cm$^2$  and $I_2\simeq 2.68 \times  10^{50}$ g
cm$^2$.   Finally  the  radii   of  the  isolated  white  dwarfs  are,
respectively,  $R_1\simeq  0.009\,  R_{\sun}$ and  $R_2\simeq  0.011\,
R_{\sun}$.


\section{Initial conditions}

We have  fixed the  initial distance between  the stars along  the $x$
axis, $x_{\rm  ini}$, and  their angular velocity,  $\omega$, allowing
the  initial distance  along  the  $y$ axis,  $y_{\rm  ini}$, and  the
initial  velocity   of  each  of  the   stars  $\vec{v}_i=(\pm  v_{\rm
ini},0,0)$ to be our free parameters.  Note that the relative velocity
of the intervening  white dwarfs is thus $2v_{\rm  ini}$.  The initial
distance  at which  the two  interacting  white dwarfs  are placed  is
always $x_{\rm  ini}=0.2 \, R_{\sun}$,  which is much larger  than the
radii of the intervening  white dwarfs ($\sim 0.01\,R_{\sun}$).  Under
these  conditions the  tidal  deformations of  both  white dwarfs  are
negligible at the beginning of the simulation and the approximation of
spherical symmetry is valid.  Note  that with this setting the initial
coordinates of  both stars are  $(+x_{\rm ini}/2, -y_{\rm  ini}/2, 0)$
and  $(-x_{\rm ini}/2,  +y_{\rm ini}/2,  0)$, and  the center  of mass
moves with  a total  velocity $v_{\rm cm}=v_{\rm  ini}/7$ in  the $xy$
plane.  The  two interacting  white dwarfs have  rotational velocities
$\omega\simeq  7\times10^{-5}$   rad/s  and  are   assumed  to  rotate
counterclockwise.  These velocities  are representative of those found
in field  white dwarfs  (Berger et al.   2005).  We have  also assumed
that the white dwarfs rotate as rigid solids (Charpinet et al. 2009).

We have  conducted 12 simulations with initial  distances ranging from
$0.1\, R_{\sun}$ to  $0.8 \, R_{\sun}$ and initial  velocities from 50
to 225 km/s  --- see Table \ref{outcome} for a  summary of the initial
conditions adopted for each of the simulations presented here.  It has
to be noted that we  have restricted our attention to the post-capture
scenario. Consequently,  all the systems studied here  were bound from
the  start.   For  a  detailed  study  of  the  gravitational  capture
mechanisms  see, for  example, Press  \& Teukolsky  (1977) and  Lee \&
Ostriker (1986). We  note, however, that in order for  a pair of stars
to  become bound  after a  close encounter,  some kind  of dissipation
mechanism must  be involved.  Some examples of  dissipation mechanisms
are  a third  body tidal  interaction (Shara  \& Hurley  2002)  or the
excitation of stellar pulsations by means of tidal interaction (Fabian
et al. 1975).

The typical  stellar dispersion velocity in  globular clusters $v_{\rm
d}$ is  approximately 10~km/s while the relative  velocity $v_{\rm c}$
in a  close encounter ---  assuming an interaction  distance $r_{\min}
\lesssim  1 R_{\sun}$  --- for  a pair  of stars  can be  shown  to be
(Fabian et al. 1975)

\begin{equation}
v_{\rm c} \simeq \left( 2G\frac{M_1+M_2}{r_{\min}}\right)^{1/2} 
\gtrsim 100~{\rm km/s}
\end{equation}

\noindent  where $M_1$  and $M_2$  are the  masses of  the interacting
stars and $r_{\min}$ is the  distance at periastron.  Thus, only $\sim
\left(  v_{\rm d}/v_{\rm c}  \right)^2 \lesssim  0.01$ of  the kinetic
energy available at  closest approach needs to be  dissipated in order
to  bound the  system.   This fraction  is  small enough  to expect  a
relatively  high  formation rate  of  this  type  of systems  (Lee  \&
Ostriker 1986). However, given our initial conditions, which result in
eccentricities  $\varepsilon  \sim  0.9$   (a  typical  value  in  the
simulations  presented  here,  see  table \ref{outcome})  more  energy
dissipation than it  is reasonable to expect from  a single periastron
passage is needed.   Hence, at least some of  the collisions presented
here are more likely to result from encounters involving three or more
stars  than they  are from  a  traditional tidal  capture (Ivanova  et
al. 2006).


\section{Results}

\begin{figure}
\begin{center}
\includegraphics[width=7cm,angle=-90]{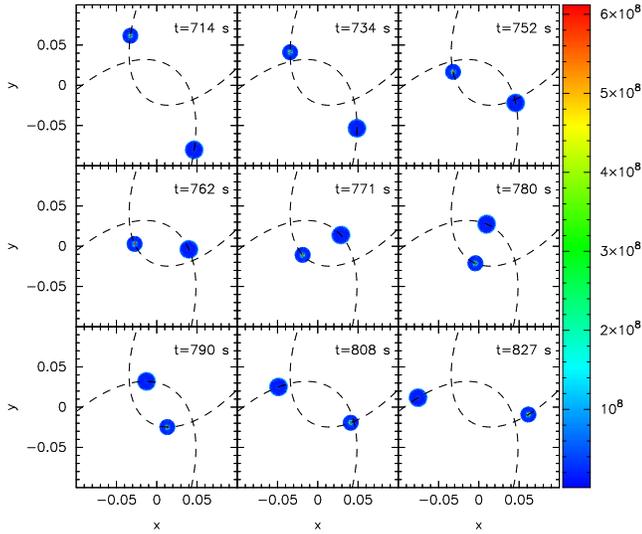}
\caption{Time  evolution  of  one  of  the simulations  in  which  the
         eccentric   double  white   dwarf  system   survives  closest
         approach.  In particular,  this simulation corresponds to the
         case in which the  initial velocity is $v_{\rm ini}=150$~km/s
         and the initial distance  is $y_{\rm ini}=0.5 R_{\sun}$.  The
         temperatures of  each SPH particle are  also shown, expressed
         in  K, while  the $x$  and  $y$ axes  are in  units of  solar
         radius.  The  dashed lines correspond to  the trajectories of
         the center of  mass of each star. Only 1  out of 10 particles
         has  been represented.  Times  are shown  in the  right upper
         corner of each panel.  These figures have been done using the
         visualization  tool SPLASH (Price  2007). [Color  figure only
         available in the electronic version of the article].}
\label{orbital}
\end{center}
\end{figure}

Depending on  the kinetic energy  dissipated during the  encounter two
different outcomes  might result: if the stars  get sufficiently close
at periastron and  mass transfer begins, a stellar  merger will occur,
otherwise  the  eccentric  binary  system  will  survive  the  closest
approach.  We have obtained three distinct behaviors, depending on the
input parameters  of the  close encounter: direct  collisions, lateral
collisions ---  characterized by  more than one  mass-transfer episode
previous  to the  stellar merger  --- and,  finally,  eccentric binary
systems surviving the closest approach.  Three representative examples
of  each   of  these  cases   are  shown,  respectively,   in  figures
\ref{orbital}, \ref{lateral} and \ref{direct}.

\subsection{Dynamics and outcomes of the interactions}

\begin{figure}
\begin{center}
\includegraphics[width=7cm,angle=-90]{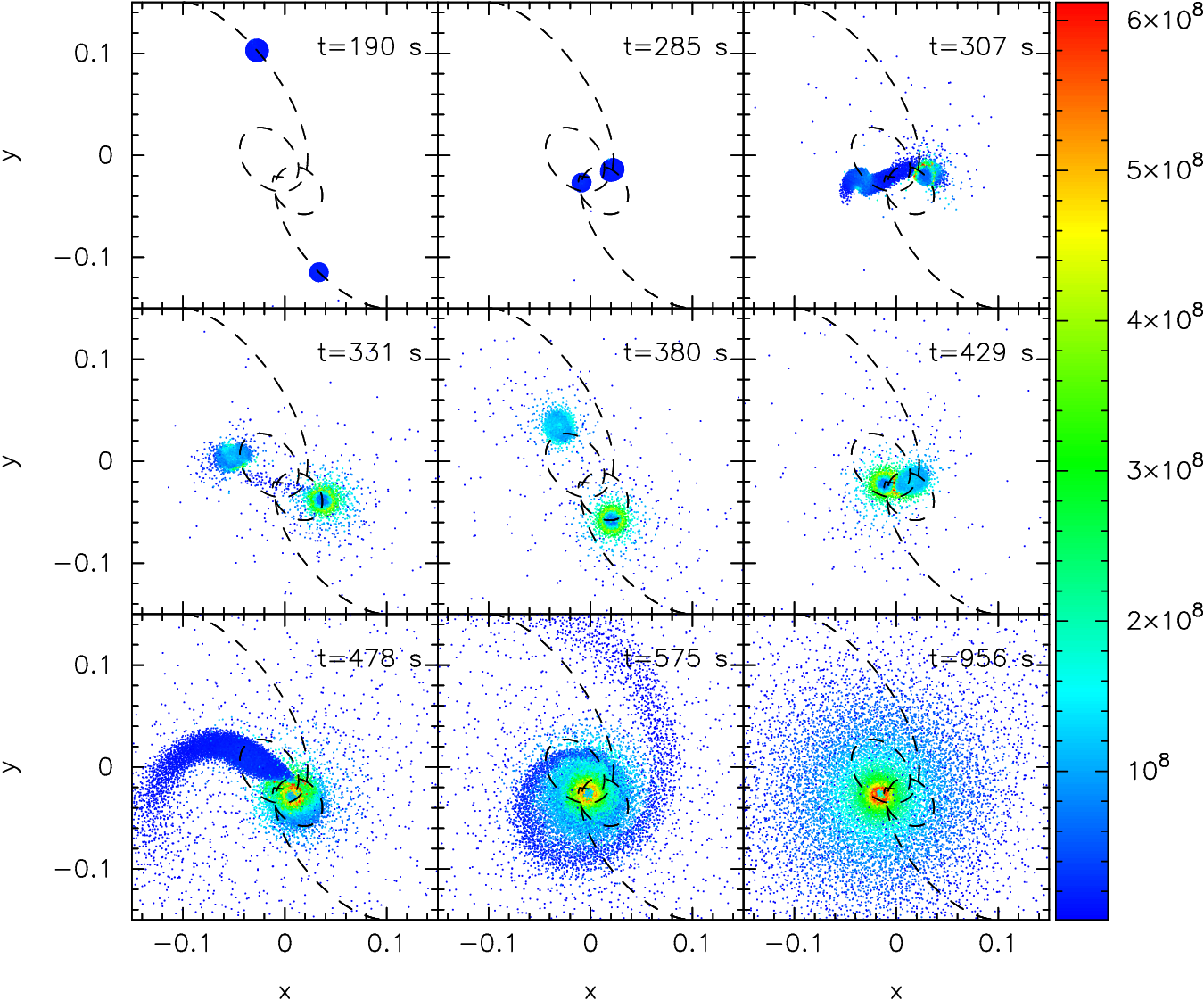}
\caption{Same as  figure \ref{orbital} for  one of the  simulations in
         which  the outcome  is a  lateral collision.   In particular,
         this corresponds to the  case in which $v_{\rm ini}=175$~km/s
         and $y_{\rm ini}=0.3 R_{\sun}$.  [Color figure only available
         in the electronic version of the article].}
\label{lateral}
\end{center}
\end{figure}

\begin{figure}
\begin{center}
\includegraphics[width=7cm,angle=-90]{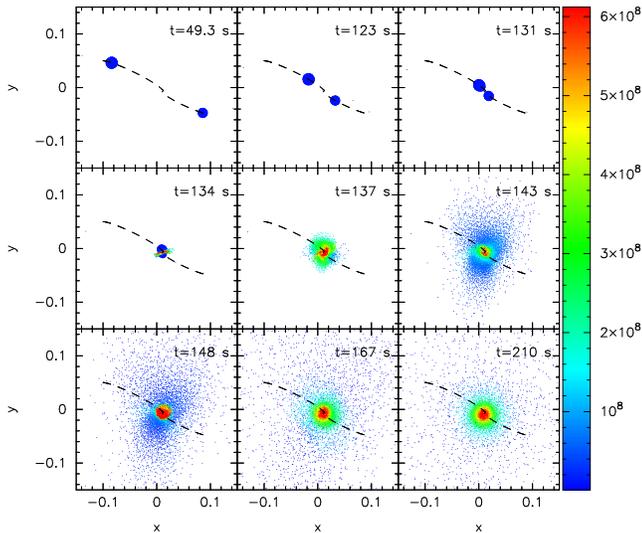}
\caption{Same as  figure \ref{orbital} for  one of the  simulations in
         which the outcome is  a direct collision. In particular, this
         corresponds to the case  for which $v_{\rm ini}=120$~km/s and
         $y_{\rm ini}=0.1 R_{\sun}$.   [Color figure only available in
         the electronic version of the article].}
\label{direct}
\end{center}
\end{figure}

In figure \ref{orbital} we show the time evolution of one of the cases
in  which the  eccentric  binary survives  the  closest approach.   In
particular, this  figure corresponds  to the case  in which a  pair of
white dwarfs of masses $0.6$ and $0.8~M_{\sun}$ interacts with initial
parameters $v_{\rm ini}=150$~km/s and $y_{\rm ini}=0.5~R_{\sun}$.  The
dashed line depicts  the motion of the center of  masses of each white
dwarf.  As can be seen, the orbits are elliptical and the temperatures
of  the   intervening  white   dwarfs  remain  constant.    In  figure
\ref{lateral} we show an example of a lateral collision, corresponding
to the same  intervening white dwarfs, but now  the initial conditions
are  $v_{\rm ini}=175$~km/s and  $y_{\rm ini}=0.3~R_{\sun}$.   In this
case the two interacting white  dwarfs perform a first passage through
the periastron  during which a first mass-transfer  episode occurs ---
see  the  top  three  panels  of  figure  \ref{lateral}.   After  this
mass-transfer episode  the two white  dwarfs become detached  --- left
and middle central panels  of figure \ref{lateral}.  Subsequently, the
two white dwarfs approach again  each other and a second mass-transfer
episode ensues --- right  central panel of figure \ref{lateral}.  This
second  mass-transfer  episode  is   unstable  and  the  white  dwarfs
coalesce. The  less massive white dwarf  forms an spiral  arm --- left
bottom  panel ---  which,  as  a consequence  of  the orbital  motion,
entangles  --- middle  bottom panel  --- and,  finally, forms  a heavy
keplerian disk --- right bottom panel --- very much in the same manner
as it occurs  in the case of coalescing  binaries (Lor\'en--Aguilar et
al.  2009). Finally, in figure \ref{direct} we display an example of a
direct collision, corresponding to the  case in which two white dwarfs
of  masses $0.6$  and $0.8~M_{\sun}$  with initial  parameters $v_{\rm
ini}=120$~km/s and $y_{\rm ini}=0.1~R_{\sun}$ collide.  As can be seen
in this  figure, during the first  stages of the  encounter both white
dwarfs preserve their original  spherical shape and their temperatures
remain stable --- top panels of figure \ref{direct}.  At $t\simeq 134$
s ---  left central  panel ---  the two white  dwarfs collide  and the
material increases considerably the temperature, reaching temperatures
as high as $T\sim 9\times 10^8$.  As a result of the direct collision,
the  SPH particles  acquire very  large velocities  ---  right central
panel  of figure  \ref{direct}  ---  and the  cloud  of SPH  particles
expands.  Initially, the  expansion of this cloud of  particles is not
perfectly symmetric  --- left bottom panel of  figure \ref{direct} ---
but at time  passes by, a spherically symmetric  cloud forms --- right
bottom panel of figure \ref{direct}.

\begin{table*}
\begin{center}
\begin{tabular}{cccccccccccc}
\hline
\hline
\noalign{\smallskip}
 Run & $M_{\rm WD}$
     & $M_{\rm debris}$
     & $M_{\rm ej}$
     & $T_{\max}$
     & $T_{\rm peak}$
     & $R_{\rm debris}$
     & $\Delta t$
     & $E_{\rm nuc}$
     & $E_\nu$
     & $E_{\rm GW}$\\
\noalign{\smallskip}
\hline
\noalign{\smallskip}
3  & 1.00 & 0.40 & $1.6\times10^{-2}$ & $5.9\times10^8$ & $1.6\times10^9$ & 0.2 & 150  & $2.5\times10^{44}$ & $3.6\times10^{30}$ & $4.2\times10^{40}$ \\
5  & 0.94 & 0.46 & $4.4\times10^{-2}$ & $5.1\times10^8$ & $6.8\times10^8$ & 0.2 & 2200 & $2.0\times10^{30}$ & $1.4\times10^{22}$ & $1.0\times10^{41}$ \\
6  & 0.88 & 0.52 & $2.2\times10^{-2}$ & $5.1\times10^8$ & $1.6\times10^9$ & 0.2 & 400  & $5.8\times10^{41}$ & $3.2\times10^{29}$ & $6.2\times10^{40}$ \\
7  & 0.91 & 0.49 & $1.3\times10^{-2}$ & $4.7\times10^8$ & $9.6\times10^8$ & 0.2 & 200  & $7.6\times10^{37}$ & $1.9\times10^{26}$ & $4.1\times10^{40}$ \\
8  & 0.84 & 0.56 & $2.7\times10^{-2}$ & $5.5\times10^8$ & $3.2\times10^9$ & 0.2 & 133  & $7.4\times10^{45}$ & $3.5\times10^{31}$ & $1.4\times10^{41}$ \\
9  & 0.82 & 0.58 & $2.5\times10^{-2}$ & $5.4\times10^8$ & $3.8\times10^9$ & 0.2 & 120  & $1.1\times10^{47}$ & $4.5\times10^{31}$ & $1.6\times10^{41}$ \\
10 & 0.67 & 0.73 & $4.6\times10^{-2}$ & $6.2\times10^8$ & $4.6\times10^9$ & 0.2 & 100  & $1.3\times10^{48}$ & $1.3\times10^{48}$ & $1.4\times10^{41}$ \\
11 & 0.73 & 0.67 & $5.3\times10^{-2}$ & $6.9\times10^8$ & $5.2\times10^9$ & 0.2 & 90   & $3.0\times10^{48}$ & $5.1\times10^{32}$ & $1.0\times10^{41}$ \\
12 & 0.78 & 0.62 & $5.2\times10^{-2}$ & $7.1\times10^8$ & $5.4\times10^9$ & 0.2 & 90   & $3.9\times10^{48}$ & $9.3\times10^{32}$ & $8.3\times10^{40}$ \\
\noalign{\smallskip}
\hline
\hline
\end{tabular}
\end{center}
\caption{Summary of  hydrodynamical results.  Masses and  radii are in
         solar  units, times  in seconds  and energies  in  ergs.  The
         maximum temperature achieved  during each simulation, $T_{\rm
         peak}$,  and  the temperature  of  the  debris region  formed
         around  the primary at  the end  of the  simulations, $T_{\rm
         max}$, are discussed in the text.}
\label{tab-hydro}
\end{table*}

In table \ref{outcome} we list,  in addition to the initial parameters
of  each  of  the  simulations   studied  here,  the  outcome  of  the
interaction. When  the interaction  of the white  dwarfs leads  to the
survival of  the eccentric binary we  label it as ``O'',  while when a
lateral  collision occurs  we use  ``LC'' and  when the  outcome  is a
direct collision we label the simulation as ``DC''. Table 1 also lists
for each simulation the total energy of the system, $E$, and its total
angular  momentum,  $L$.  Since  all  the  systems  studied here  have
negative      energies     their     initial      trajectories     are
elliptical. Consequently,  we also  list the perigee  ($r_{\min}$) and
the  apogee  ($r_{\max}$)  of  the  initial  orbit,  and  the  initial
eccentricity.   These   distances  have  been   calculated  using  the
well-known  solution of the  two-body problem,  assuming that  the two
white  dwarfs  are  point  masses.   Namely,  we  use  $\varepsilon  =
\sqrt{1+(2EL^2)/(\mu  k^2)}$,  where  $k=GM_1M_2$  and  $\mu$  is  the
reduced    mass   of    the   system.     Finally,   we    also   list
$\beta=(R_1+R_2)/r_{\min}$, where $R_1$ and $R_2$ are the radii of the
interacting  white dwarfs, which  is a  parameter which  describes the
strength  of  the encounter.   Large  values  of  $\beta$ imply  large
interaction strengths.

\begin{figure*}
\begin{center}
\vspace{8.0cm}  
\includegraphics{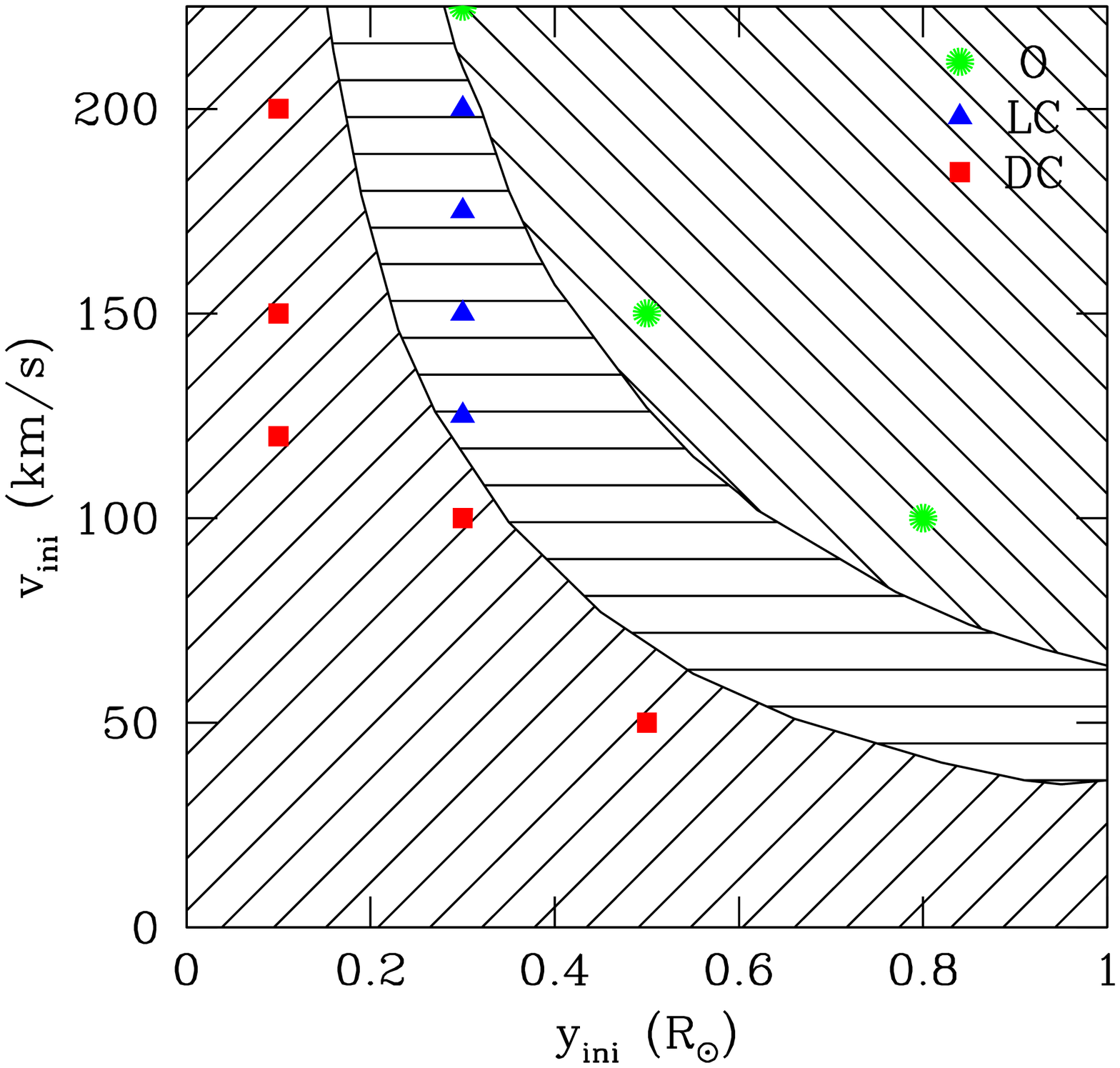}
\includegraphics{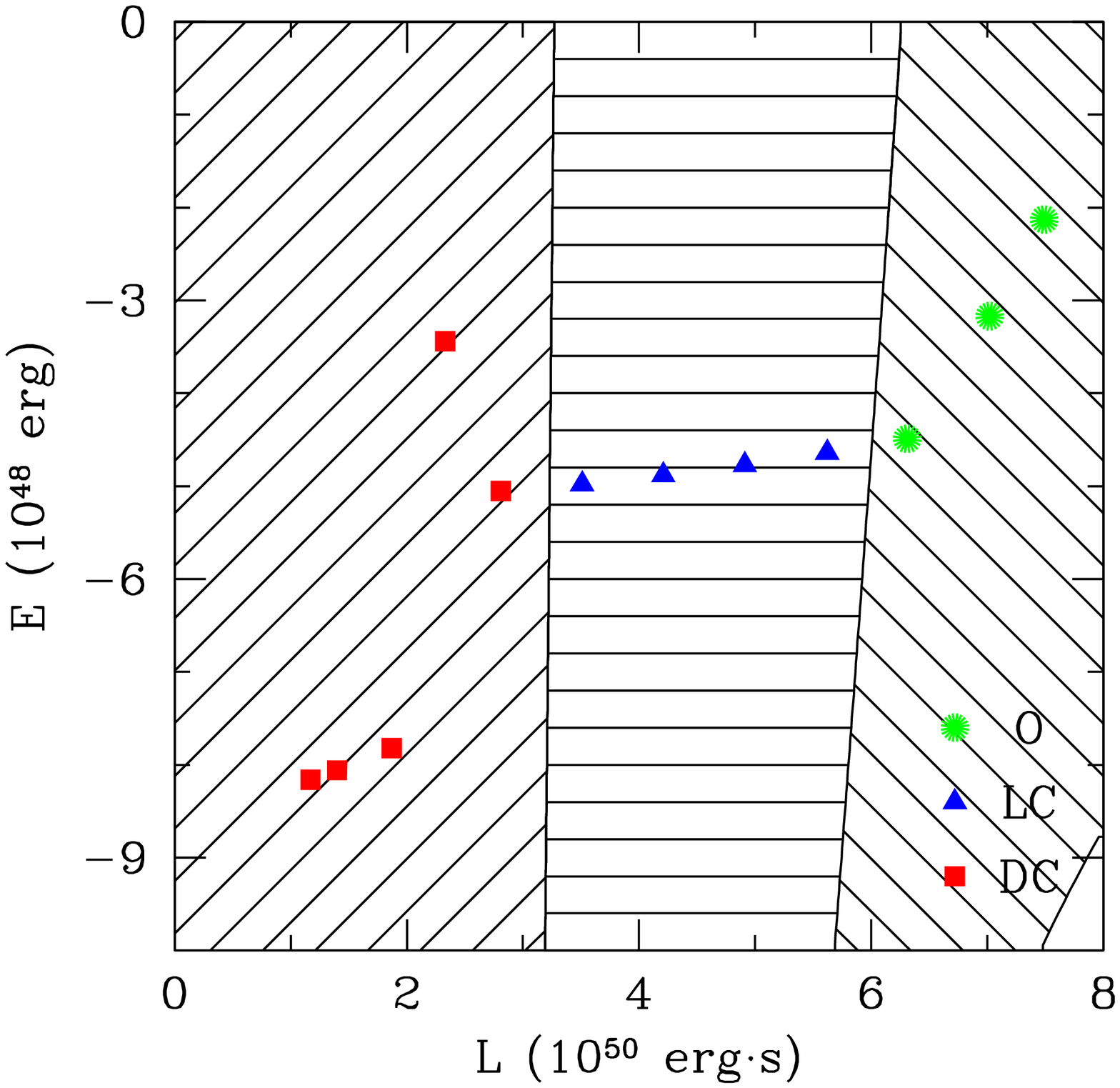}
\caption{Left panel: regions  in the initial velocity-initial distance
         plane  for which the  three different  outcomes of  the white
         dwarf close encounters occur  for the $0.6\, M_{\sun} + 0.8\,
         M_{\sun}$ interaction studied  here. The simulations in which
         a direct collision (DC) occurs are shown using solid squares,
         solid  triangles show  the  simulations for  which a  lateral
         collision  (LC)  occurs, whilst  those  in  which the  binary
         system survives (O) are represented using solid circles.  For
         sufficiently  large initial  distancess  an eccentric  binary
         system results,  while for  small initial distances  a direct
         collision happens. In the  left and middle shaded regions the
         outcome is a merger.  Right  panel: same as the left panel in
         the  energy-angular momentum  plane.  We  emphasize  that the
         transitions between the  different regions have been computed
         using  Eqs.  (7)  to  (9),  which are  valid  for  point-like
         masses.  [Color  figure  only  available  in  the  electronic
         version of the article].}
\label{regions}
\end{center}
\end{figure*}

A first  look at  table \ref{outcome} reveals  that the  most relevant
parameter  for  discriminating between  the  three different  outcomes
previously  discussed is  the periastron  distance, $r_{\min}$,  as it
should be expected.  If the distance at the periastron is smaller than
$\simeq  0.033\pm 0.004~R_{\sun}$  mass transfer  begins and  either a
lateral collision or a direct one occurs and, thus, a merger turns out
to be the unavoidable outcome.   However, depending on the exact value
of $r_{\min}$ the merger occurs  in two or more mass-transfer episodes
--- for  $0.009\pm   0.002~R_{\sun}  \lesssim  r_{\rm   min}  \lesssim
0.033\pm 0.004~R_{\sun}$  --- or by  means of a direct  collision, for
$r_{\rm min} \lesssim 0.009\pm 0.002~R_{\sun}$.

In  figure  \ref{regions}  we  show  the  different  outcomes  of  the
interactions  as a function  of the  initial velocities  and distances
(left  panel)  and as  a  function of  the  total  energy and  angular
momentum of each simulation  (right panel).  Red squares correspond to
simulations in  which the  final outcome is  a direct  collision, blue
triangles  to  those in  which  we  obtain  a lateral  collision  and,
finally, green  circles to those  in which an eccentric  binary system
survives.  Finally, the theoretical combinations of initial parameters
that  lead to the  three different  outcomes previously  discussed are
represented  in  these planes  using  different  shaded regions.   The
borders  of  these regions  have  been  obtained  using the well-known
solution of the two body problem for the periastron distance
\begin{equation}
 r_{\min} = \frac{k}{2\mid E \mid}\left(1-\sqrt{1+\frac{2EL^2}{\mu k^2}}\right)
\end{equation}
and  imposing, respectively,  $r_{\min}=0.033$  and $0.009\,R_{\sun}$.
The  regions of  the  left  panel of  figure  \ref{regions} have  been
obtained using the same procedure and taking into account that
\begin{eqnarray}
r_{12} &=& \left(x_{\rm ini}^2 + y_{\rm ini}^2\right)^{1/2} \\
E &=& \frac{1}{2}\left(M_1+M_2\right)v_{\rm ini}^2
    - \frac{GM_1M_2}{r_{12}}
  -\frac{1}{2}\left(M_1+M_2\right){v_{\rm cm}}^2 \nonumber\\
  &=&2\mu v_{\rm ini}^2  - \frac{G(M_1+M_2)\mu}{r_{12}}\\
L &=& \frac{1}{2}\left(M_1+M_2\right)v_{\rm ini} y_{\rm ini} \nonumber\\
  &+& \left(M_1+M_2\right)\left(x_{\rm cm}v_{{\rm cm}_y} - 
      y_{\rm cm}v_{{\rm cm}_x}\right)\nonumber\\
  &=& 2\mu y_{\rm ini} v_{\rm ini}
\end{eqnarray}
As  can be  seen,  for sufficiently  large  initial $y$-distances  the
eccentric  binary system  survives  the interaction,  while for  small
initial  distances a  direct  collision always  occurs.  Between  both
regions there exists another one in which a lateral collision occurs.

In all the cases in  which a merger occurs the resulting configuration
consists of a central white  dwarf and a debris region. The morphology
of this  region will be discussed below.   Table \ref{tab-hydro} lists
several important physical quantities.  Specifically, for the cases in
which a lateral or a direct  collision occurs --- and, thus, for those
cases in which the final result  of the interaction is a central white
dwarf surrounded by  the debris of the collision ---  we list the mass
of the central  white dwarf ($M_{\rm WD}$) obtained at  the end of the
interaction,  the  mass  of  the  debris  ($M_{\rm  debris}$)  of  the
collision, and the mass ejected during the interaction ($M_{\rm ej}$).
A precise definition of these masses is given in \S 4.2. Additionally,
we  list the peak  temperature achieved  for each  simulation, $T_{\rm
peak}$ which  will be useful when discussing  the chemical composition
of  the remnants,  the  maximum temperature,  $T_{\max}$,  of the  hot
region that we  find on top of the primary white  dwarf, the radius of
the  region where  the debris  of  the collisions  are found,  $R_{\rm
debris}$, the  total time in  which the collision occurs,  $\Delta t$,
and the  nuclear, neutrino and gravitational  energies released during
the   interaction   ($E_{\rm  nuc}$,   $E_\nu$,   and  $E_{\rm   GW}$,
respectively).  The duration  of the collision is defined  as the time
elapsed  since  mass  transfer  begins  until  the  system  reaches  a
symmetric  configuration.  All  these quantities  are of  interest for
discussing the structure and nucleosynthesis of the merger remnants.

It is worth noting that for runs 3 to 7 a good fraction of the mass of
the less massive  white dwarf is accreted onto  the more massive white
dwarf, thus increasing the total  mass of the central remnant.  In all
these cases  the collisions  are rather gentle  and thus  although the
interactions  result in a  merger and  occur in  dynamical timescales,
mass transfer  happens during relatively  long times (see column  8 of
table  \ref{tab-hydro}).  In  the case  of runs  number 8  and  9 very
little mass is  accreted by the more massive  white dwarf, whereas for
runs 10,11  and 12 the collision is  so strong that the  impact of the
less massive white  dwarf removes mass from the  more massive one and,
consequently, the mass  of the mass of the  central remnant is smaller
than that of the original massive  white dwarf.  Note as well that the
mass ejected  during the  interaction is relatively  small in  all the
cases, of the order of  $\sim 10^{-2}\, M_{\sun}$, so the interactions
are almost conservative. Consequently,  the mass of debris surrounding
the  central  merged  object  follow  a trend  reverse  of  that  that
described before  for the central  white dwarf.  This behavior  can be
related to  the previously  defined interaction strength  $\beta$ (see
table \ref{outcome}). For runs 3, 4,  5, 6 and 7 we have $\beta=4.00$,
0.54, 0.69  and 0.95, respectively, whilst  for runs 10, 11  and 12 we
obtain  6.67,  10.0  and  20.0.   Clearly,  the  former  runs  can  be
considered as  relatively mild, whereas  the later are  rather strong.
However, it is interesting to note that, at odds with what occurs with
the debris masses, the size of the regions in which we find the debris
of  the   collisions  is   totally  independent  of   the  kinematical
characteristics  of the  interacting  white dwarf  ($y_{\rm ini}$  and
$v_{\rm ini}$).  We emphasize,  however, that the density distribution
is  very different  depends very  much on  these  characteristics (see
below).

\subsection{Structure of the merger remnants}

\begin{figure}
\begin{center}
\vspace{8.2cm}  
\includegraphics{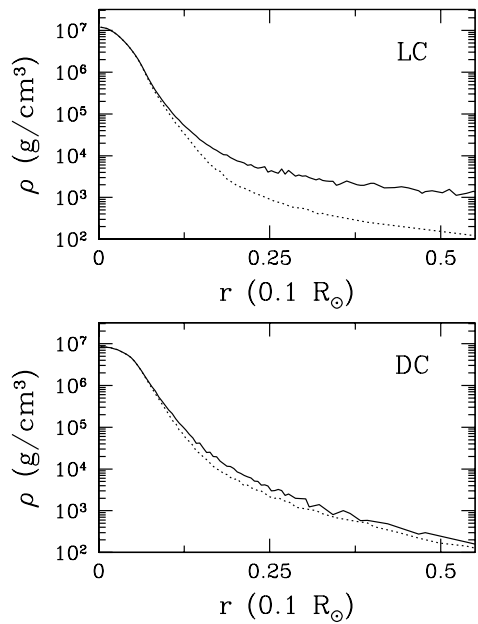}
\caption{Density profiles  of the merged configurations at  the end of
         our  simulations for  the white  dwarf interactions  shown in
         figures  \ref{lateral} (top  panel) and  \ref{direct} (bottom
         panel).  The solid lines  correspond to  the profiles  in the
         equatorial plane, whereas the dashed line are the profiles in
         the polar direction.}
\label{Dfinal}
\end{center}
\end{figure}

Due  to the  different dynamics  of the  white dwarf  interactions the
resulting merged configurations are  not the same.  To illustrate this
in  Fig. \ref{Dfinal}  we  plot  the density  profiles  for a  lateral
collision (top  panel) and  a direct collision  (bottom panel)  in two
directions. The  solid line depicts  the profile along  the equatorial
plane and the  dashed line in the polar direction.  As  can be seen in
figure \ref{Dfinal},  in the case of  a lateral collision  we obtain a
completely  different merged  configuration  than that  obtained in  a
direct  collision.   Specifically,  the  density  profiles  along  the
equatorial plane and  along the polar direction show  that in the case
of  a lateral  collision the  result of  the interaction  is  a nearly
spherically symmetrical compact object surrounded by an extended disk,
whilst in the case of a direct collision the result of the interaction
is a central compact object surrounded by a nearly spherical cloud.

\begin{figure}
\begin{center}
\vspace{8.2cm}  
\includegraphics{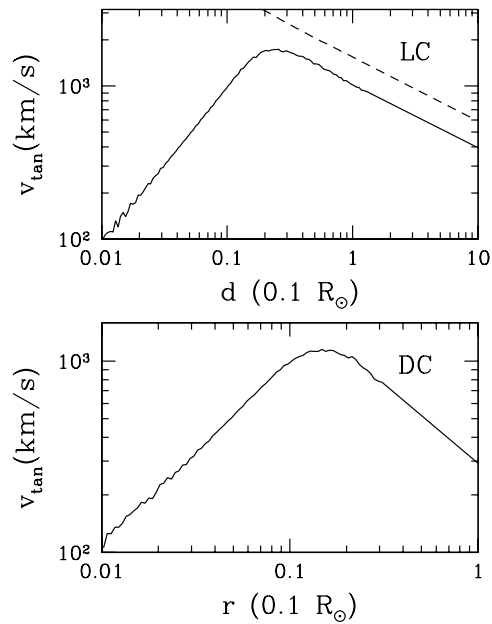}
\caption{Distributions of the angular  velocity --- solid lines --- at
         the end  of our simulations for the  white dwarf interactions
         shown in  figures \ref{lateral} (top  panel) and \ref{direct}
         (bottom panel). See main text for details.}
\label{vfinal}
\end{center}
\end{figure}

Figure  \ref{vfinal} shows  the  rotational velocities  of the  merger
remnants for the case of a  lateral collision (top panel) and a direct
collision (bottom  panel).  For  the case of  a lateral  collision the
rotational  velocities are plotted  as a  function of  the cylindrical
radius and we have averaged the velocities in concentrical cylindrical
shells,  whereas for  the  case of  a  direct collision  we have  used
instead spherical coordinates.  For the case of a lateral collision we
also show a  keplerian profile, using a dashed  line. In these figures
we have adopted a reference  system comoving with the central remnant.
To compute the  position and velocity of this  reference frame we have
computed the center  of mass of the remnant, as  well as its velocity.
To this  end we  only considered those  particles of the  remnant with
densities  larger  than a  certain  threshold,  which  we have  chosen
$\rho=6.0\times10^3$ g/cm$^3$.   As can  be seen in  the top  panel of
Fig.  \ref{vfinal}, corresponding to  a lateral collision, the remnant
is made  of a  central compact object  (the most massive  white dwarf)
which rotates as  a rigid solid and, on  top of it, it can  be found a
keplerian disk, which is the product of the entanglement of the spiral
arm which forms from the disrupted less massive white dwarf. We remind
that this entanglement occurs  during the second mass-transfer episode
--- see  Fig.   \ref{lateral}.  Finally,  for  the  case  of a  direct
collision  we also  obtain  that  the central  region  of the  remnant
rotates with  constant angular velocity.  Additionally, on  top of the
central object we  find a nearly spherically symmetric  cloud in which
the tangential  velocity has a  profile with a  $r^{-5/3}$ dependance.
This morphology is  a direct consequence of the  large strength of the
collision.  In fact, we find that  the collision is so strong that the
particles  of the colliding  white dwarfs  are well  mixed in  all the
regions of the remnant. However, a cautionary remark is in order here.
In particular, the rigid rotation of the merged configurations seen in
this figure  is indicative that,  even though the  numerical viscosity
may be  small, our artificial  viscosity prescription is  still larger
than  the correct  physical viscosity.   Consequently, it  is possible
that  the rigid  rotation found  in our  simulations is  not physical,
since  it  is hard  to  avoid  spurious  effects from  the  artificial
viscosity for runs that last many dynamical timescales.

The  masses of  these two  regions for  all the  simulations presented
here, as  well as the  mass ejected in  each run, are listed  in Table
\ref{tab-hydro}. According to the  previous discussion the mass of the
central  white dwarf  corresponds to  the mass  of the  region  of the
remnant which has  constant angular velocity.  The mass  of the ejecta
simply  corresponds  to  the   mass  of  those  particles  which  have
velocities  larger than  the  local escape  velocity  in the  comoving
frame, while the mass of the debris is that of the region in which the
particles have either  keplerian velocities (in the case  of a lateral
collision) or  which rotate with  a profile with slope  $r^{-5/3}$ (in
the case of a direct collision).

\subsection{Temperatures and nucleosynthesis}

\begin{table*}
\centering
\begin{tabular}{ccccccccc}
\hline
\hline
\noalign{\smallskip}
Run & He & C & O & Ne & Mg & Si & S & Ar \\
\noalign{\smallskip}
\hline
\noalign{\smallskip}
3  & $4\times10^{-9} $ & 0.4   & 0.6   & $5\times10^{-5}$  & $6\times10^{-8}$  & $3\times10^{-11}$ & 0                 & 0                 \\
5  & 0                 & 0.4   & 0.6   & 0                 & 0                 & 0                 & 0                 & 0                 \\
6  & $1\times10^{-10}$ & 0.4   & 0.6   & $2\times10^{-7}$  & $6\times10^{-10}$ & $2\times10^{-13}$ & 0                 & 0                 \\
7  & $5\times10^{-13}$ & 0.4   & 0.6   & $7\times10^{-12}$ & 0                 & 0                 & 0                 & 0                 \\
8  & $1\times10^{-9}$  & 0.4   & 0.6   & $4\times10^{-5}$  & $6\times10^{-6}$  & $1\times10^{-7}$  & $1\times10^{-10}$ & 0                 \\
9  & $2\times10^{-8}$  & 0.399 & 0.599 & $7\times10^{-4}$  & $8\times10^{-5}$  & $1\times10^{-6}$  & $2\times10^{-9}$  & $2\times10^{-13}$ \\
10 & $6\times10^{-8}$  & 0.398 & 0.599 & $3\times10^{-3}$  & $3\times10^{-4}$  & $4\times10^{-6}$  & $5\times10^{-9}$  & $6\times10^{-13}$ \\
11 & $1\times10^{-7}$  & 0.397 & 0.598 & $3\times10^{-3}$  & $3\times10^{-4}$  & $6\times10^{-6}$  & $7\times10^{-9}$  & $1\times10^{-12}$ \\
12 & $2\times10^{-7}$  & 0.397 & 0.598 & $4\times10^{-3}$  & $4\times10^{-4}$  & $7\times10^{-6}$  & $8\times10^{-9}$  & $1\times10^{-12}$ \\
\hline
\hline
\end{tabular}
\caption{Averaged chemical  composition (mass fractions)  of the heavy
         rotationally-supported   disk  (in   the   case  of   lateral
         collisions)  or  cloud (in  the  case  of direct  collisions)
         obtained  by the  end of  the interaction,  for the  cases in
         which a merger is the outcome.}
\label{tab-chem-disk}
\end{table*}

\begin{figure}
\begin{center}
\vspace{8.2cm}  
\includegraphics{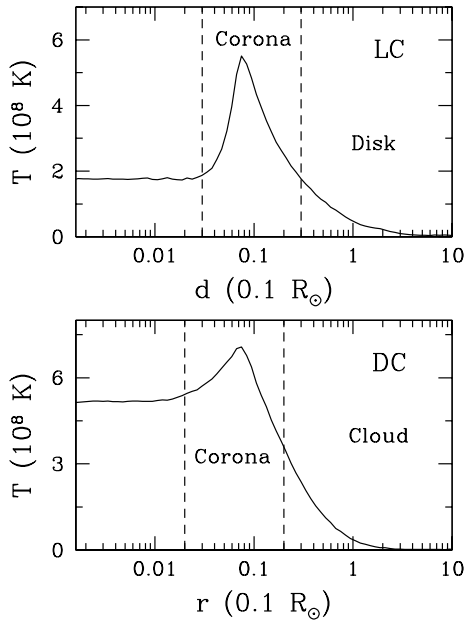}
\caption{Temperature profiles of the  merged configurations at the end
         of our simulations for  the white dwarf interactions shown in
         figures  \ref{lateral} (top  panel) and  \ref{direct} (bottom
         panel). In  the case of  a lateral collision the  abscissa is
         the cylindrical  radius, $d$, while  in the case of  a direct
         collision we use the spherical radius.}
\label{Tfinal}
\end{center}
\end{figure}

In figure  \ref{Tfinal} we display the final  temperature profiles for
our fiducial  cases in which a  lateral and a  direct collision occur,
upper  and lower panel  respectively.  These  profiles prove  that the
maximum  temperatures occur  very close  to  the edge  of the  central
coalesced white dwarf,  in the rapidly rotating regions  which we have
previously  described, and  that the  maximum temperatures  are rather
high, in excess of $T\sim 5\times  10^8$ K. Note as well that in these
figures, as  it was done in  Fig.  \ref{vfinal}, we  use a logarithmic
scale to better display the regions  of interest.  As can be seen, the
central regions of the  merger product are practically isothermal, and
on  top of  the  relatively cold,  degenerate  core a  region of  high
temperatures is present.   We refer to this region  as the hot corona.
On top  of this region  we find  the keplerian disk  in the case  of a
lateral collision and a hot spherically symmetric cloud in the case of
a direct collision. The boundaries of the hot corona are clearly shown
in figure \ref{Tfinal} using dashed lines.  We also emphasize that for
the case of a direct collision, the temperature of the isothermal core
is almost  three times larger  than that obtained  in the case  of the
lateral  collision, a  direct  consequence of  the larger  interaction
strength.

In column 5  of table \ref{tab-hydro} we show  the maximum temperature
when the merger process has finished, for those cases in which this is
the  outcome  of  the  simulation.    As  can  be  seen,  the  maximum
temperatures of the hot coronae or cloud previously described increase
as the interaction strength ($\beta$) increases, as one should expect,
given that in  these cases more mechanical energy  is transformed into
thermal energy.  Additionally, in column 6 of table \ref{tab-hydro} we
also list  the peak temperature  --- that is, the  maximum temperature
--- achieved during  the entire simulation, $T_{\rm  peak}$, for those
cases  in  which a  merger  is  obtained  after the  interaction.   As
expected,  the peak  temperature increases  with the  strength  of the
interaction  and,  consequently,  also  do the  nuclear  and  neutrino
energies released during the interaction --- columns 9 and 10 of table
\ref{tab-hydro}. These peak temperatures occur during the most violent
phases  of  the close  encounter,  when  mass-transfer  from the  less
massive white dwarf takes place at  very high rates.  In all the cases
studied here the  peak temperature is in excess  of $T_{\rm peak} \sim
1.6  \times   10^9$  K,  clearly  higher  than   the  carbon  ignition
temperature $T_{\rm ign}\sim 10^9$ K, and leads to significant nuclear
processing. However, a strong  thermonuclear flash does not develop in
any  of these  simulations because,  although the  temperature  in the
region where the  material of the less massive  white dwarf first hits
the  more massive one  increases very  rapidly, degeneracy  is rapidly
lifted,  leading to  an expansion  of  the material,  which, in  turn,
quenches  the thermonuclear flash.   This agrees  with the  results of
Guerrero et al.   (2004), Yoon et al.  (2007)  and Lor\'en--Aguilar et
al. (2009)  for the case  in which the  coalescence of a  double white
dwarf binary  system occurs.  Thus, since these  high temperatures are
only  attained  during  a  very  short  time  interval,  thermonuclear
processing  is relatively  mild in  all  simulations. This  is not  in
contradiction with the results of  Raskin et al. (2009) and Rosswog et
al.   (2009), as these  authors computed  collisions with  much larger
collision strengths.

\begin{table*}
\centering
\begin{tabular}{ccccccccccccccc}
\hline
\hline
\noalign{\smallskip}
Run & He & C & O & Ne & Mg & Si & S & Ar \\
\noalign{\smallskip}
\hline
\noalign{\smallskip}
3  & $1\times10^{-9}$  & 0.4   & 0.6   & $3\times10^{-5}$  & $3\times10^{-8}$ & $2\times10^{-11}$ & 0                 & 0                 \\
5  & 0                 & 0.4   & 0.6   & 0                 & 0                & 0                 & 0                 & 0                 \\
6  & $8\times10^{-14}$ & 0.4   & 0.6   & $1\times10^{-12}$ & 0                & 0                 & 0                 & 0                 \\
7  & $1\times10^{-13}$ & 0.4   & 0.6   & $2\times10^{-12}$ & 0                & 0                 & 0                 & 0                 \\
8  & $1\times10^{-11}$ & 0.4   & 0.6   & $5\times10^{-8}$  & $3\times10^{-12}$& 0                 & 0                 & 0                 \\
9  & $2\times10^{-10}$ & 0.4   & 0.6   & $7\times10^{-5}$  & $8\times10^{-6}$ & $1\times10^{-7}$  & $1\times10^{-10}$ & $1\times10^{-14}$ \\
10 & $2\times10^{-9}$  & 0.399 & 0.599 & $2\times10^{-3}$  & $2\times10^{-4}$ & $3\times10^{-6}$  & $3\times10^{-9}$  & $2\times10^{-13}$ \\
11 & $4\times10^{-9}$  & 0.398 & 0.598 & $3\times10^{-3}$  & $3\times10^{-4}$ & $4\times10^{-6}$  & $3\times10^{-9}$  & $3\times10^{-13}$ \\
12 & $5\times10^{-9}$  & 0.397 & 0.598 & $4\times10^{-3}$  & $3\times10^{-4}$ & $4\times10^{-6}$  & $4\times10^{-9}$  & $4\times10^{-13}$ \\
\noalign{\smallskip}
\hline
\hline
\end{tabular}
\caption{Averaged  chemical composition  (mass fractions)  of  the hot
         corona  obtained by  the end  of the  interaction,  for those
         cases in  which the  outcome of the  interaction is  either a
         lateral or a direct collision.}
\label{tab-chem-corona}
\end{table*}

The chemical composition of the  disk or cloud formed as a consequence
of the close encounter can  be found for all the simulations presented
in this  paper in table  \ref{tab-chem-disk}.  In this table  we show,
for  each  of  the   mergers  computed  here,  the  averaged  chemical
composition  (mass fractions)  of  the heavily  rotationally-supported
disk or cloud  described previously. We do not  show the abundances of
Ca, Ti, Cr,  Fe, Ni and Zn because they are  negligible.  As it should
be expected,  the amount of  nuclearly processed matter  increases for
increasing  interaction strenghts.  Specifically,  for the  mergers in
which in which  the interaction strength is small  ($\beta\sim 1$) the
abundance of  Ne is, at  most, of the  order $10^{-5}$ by  mass, while
those of Mg and Si are  much smaller, and we do not obtain significant
amounts of  S and Ar. Instead,  for the cases in  which $\beta\sim 10$
Ne, with a mass abundance of  $\sim 10^{-3}$ is rather abundant and we
also find significant  amounts of Mg, Si, S  and Ar.  These abundances
are  in line  with those  obtained for  the coalescence  of  two white
dwarfs in a binary system (Lor\'en--Aguilar et al. 2009).

In table  \ref{tab-chem-corona} we list  the mass abundances  of heavy
nuclei in  the hot region at the  edge of the central  white dwarf for
the same cases  listed in table \ref{tab-chem-disk}. We  find that, at
odds which what  occurs in the case of the merger  of two white dwarfs
in a binary  system, these abundances are not  very different of those
obtained  for the  disk or  cloud.  The  reason for  this is  that, as
explained previously, in the most violent interactions the material of
the  most massive  white dwarf  is removed  and incorporated  into the
debris of  the interaction.  In summary, both  the hot corona  and the
debris disk  or cloud are  enhanced in Ne  and Mg, which are  the main
products of  carbon burning.   However, a cautionary  remark regarding
the chemical compositions  of the mergers studied here  must be added.
White dwarfs are  characterized in $\sim 80\%$ of the  cases by a thin
hydrogen atmosphere  of $\sim 10^{-4}\,  M_{\sun}$ on top of  a helium
buffer of $\sim 10^{-2}\, M_{\sun}$.   In the remaining $\sim 20\%$ of
the  cases,  the hydrogen  atmosphere  is  absent.   Small amounts  of
hydrogen or helium could indeed change the nucleosynthetic patterns of
the hot corona  and debris regions in all  these cases.  Studying this
possibility is beyond  the scope of this paper  and, thus, the changes
in the  abundances associated to burning  of the helium  buffer and of
the atmospheric hydrogen layer remain to be explored.

\subsection{Comparison with other works}

In this  section we compare our  results with those  obtained in other
recent works,  namely those  of Rosswog et  al.  (2009) and  Raskin et
al. (2009) and with the  results of Lor\'en--Aguilar et al. (2009) for
the merger of  white dwarfs in binary systems.  However we remark that
these  comparisons should  be taken  with  some care,  as the  initial
conditions adopted  here are quite  different of those adopted  in the
above mentioned works.

Rosswog et  al.  (2009)  studied the interaction  of two  white dwarfs
using a variety of stellar masses.  Of the cases studied in Rosswog et
al. (2009)  the most similar  one to our  choice of masses is  that in
which the  masses of the  colliding white dwarfs are  $0.6\, M_{\sun}$
and $0.9\,  M_{\sun}$.  However,  we note that  in our case  the total
mass of the system is smaller than Chandrasekhar's mass, whilst in the
case of  Rosswog et al.  (2009)  the total mass of  the system exceeds
this  mass. Rosswog et  al.  (2009)  used parabolic  trajectories and,
consequently, the  velocities of the  colliding white dwarfs  at first
contact are  much larger than  those obtained here. Hence,  shocks are
the natural  result in their  simulations. In our case  the simulation
that best matches  the initial conditions of Rosswog  et al. (2009) is
run  10,  in  which  $v_{\rm  ini}=120$ km/s  and  $y_{\rm  ini}=0.1\,
R_{\sun}$, leading to an interaction strength $\beta=20$ --- see Table
1  ---  which   is  the  maximum  interaction  strength   of  all  our
simulations. In  both cases the  initial separation is  similar, being
$x_{\rm ini}\sim 3(R_1+R_2)$ in the case of Rosswog et al.  (2009) and
$x_{\rm ini}\sim 10(R_1+R_2)$ in  our case.  Nevertheless, and despite
the  rather  different initial  conditions,  the maximum  temperatures
obtained in both simulations  agree remarkably well.  Specifically, we
obtain a  maximum temperature  $T_{\rm peak}\simeq 5.4\times  10^9$ K,
whereas Rosswog  et al.   (2009) obtain $T_{\rm  peak}\simeq 7.9\times
10^9$ K.  While the agreement  between both sets of simulations in the
case  of the  temperatures  is rather  good,  an essential  difference
between both works  is the central density of  the resulting remnants,
we  obtain  a typical  density  $\rho\sim  10^7$  g cm$^{-3}$,  whilst
Rosswog et al. (2009) obtain  a density one order of magnitude larger,
$\rho\sim  10^8$ g  cm$^{-3}$.  As  a consequence,  the  total nuclear
energy released (and, hence, the  degree of nuclear processing) in the
case  of Rosswog  et al.  (2009),  $E_{\rm nuc}\sim  10^{50}$ erg,  is
substantially  larger  than   that  obtained  here,  $E_{\rm  nuc}\sim
10^{48}$ erg.

A comparison with the results of Raskin et al. (2009) is somewhat more
difficult, as these authors do not provide all the relevant details to
which we can compare.  In  particular, they study the direct collision
of two  identical white dwarfs  of masses $0.6\, M_{\sun}$  at various
initial  $y$-distances,  namely,  $y_{\rm  ini}=0$, $0.9$  and  $1.7\,
R_{\sun}$, placed at  an initial distance of $x_{\rm  ini}=0$, 0.5 amd
$0.9 \,  R_1$.  The  general behaviour of  the simulations is  in both
cases rather similar, and  the temperature obtained is, again, similar
in  both sets of  calculations.  Specifically,  they mention  that the
peak  temperatures achieved  during the  collisions are  in  excess of
$10^9$ K, a value similar to that obtained here.

It is also  interesting to compare our results  with those obtained in
recent simulations  of the merger  of white dwarfs in  binary systems.
As mentioned, the  most recent works are those  of Lor\'en--Aguilar et
al. (2009) and Yoon et al. (2007). Both sets of simulations yield very
similar results.  In  general we find that the  characteristics of the
collisionally  merged products  are  rather similar  to  those of  the
remnants  found   in  those  studies.  However,  there   are  as  well
significant  differences.  For  instance,  we find  that  in the  most
violent collisions  the degree of  mixing of the chemical  products is
much larger  here than in  the case of  white dwarf coalescences  in a
binary  system.   Also,  the  final  temperatures  obtained  here  are
consistent with those obtained in the works of Lor\'en--Aguilar et al.
(2009) and  Yoon  et  al.   (2007),  but  depend  sensitively  on  the
interaction strength.   In contrast in  the merger of white  dwarfs in
binary systems  the interaction proceeds  through Roche-lobe overflow,
and thus this additional degree of freedom does not exist. Finally the
structure  of  merger products  is  similar  in  the case  of  lateral
collisions,  whereas  for  direct  collisions we  obtain,  as  already
mentioned, a spherical cloud instead of a keplerian disk.

\subsection{An interesting case: multiple mass-transfer episodes}

\begin{figure}
\begin{center}
\vspace{8cm}  
\includegraphics{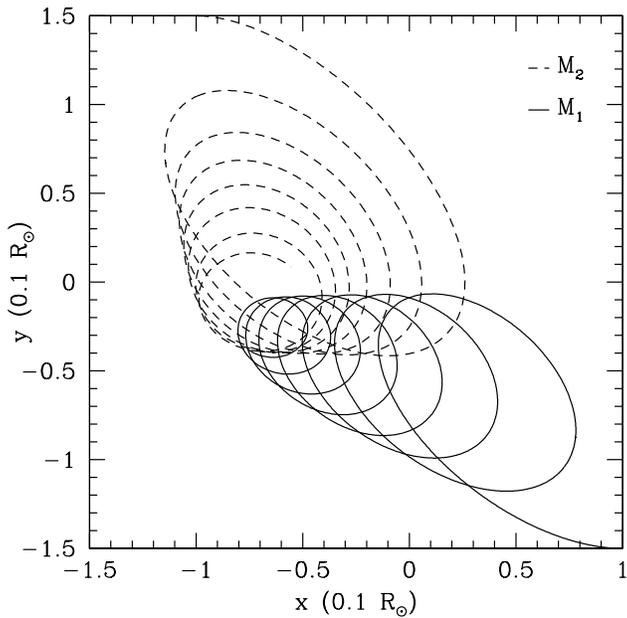}
\caption{Trajectories of the centers of mass of the most massive white
         dwarf --- solid line --- and the less massive white dwarf ---
         dashed  line   ---  for  run  number  5,   in  which  several
         mass-transfer episodes occur.}
\label{orbits}
\end{center}
\end{figure}

\begin{figure}
\begin{center}
\vspace{8cm}  
\includegraphics{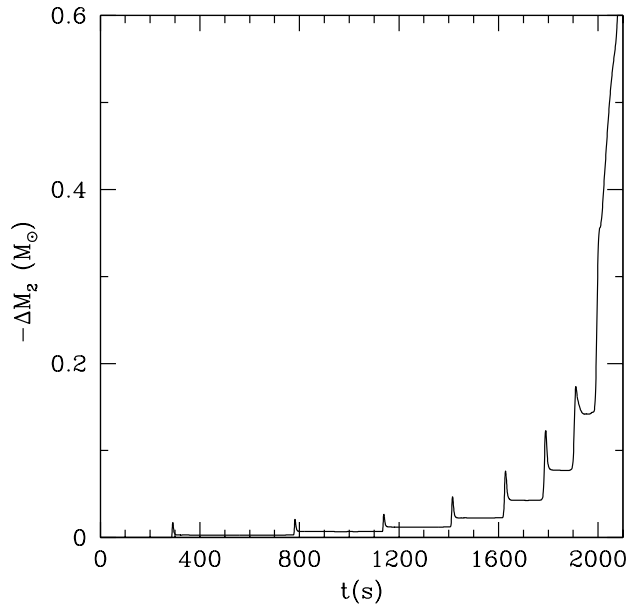}
\caption{Mass lost  by the less massive  white dwarf as  a function of
         time for run number 5.}
\label{deltam}
\end{center}
\end{figure}

Run number 5 is an interesting case and deserves further explanations.
This particular case  corresponds to a case in  which the collision is
lateral  and the  initial conditions  have  been chosen  to probe  the
transition  between a lateral  collision and  the survival  at closest
approach  of a  binary system  composed of  two white  dwarfs  --- see
figure  \ref{regions}  and  table  \ref{outcome}.   Specifically,  the
initial  conditions  are  such  that  a  very  gentle  interaction  is
obtained.   Note  that  the  interaction  strength  is  in  this  case
$\beta\sim  0.69$.   Thus, this  is  a  representative  case of  those
interactions  in  which  several  mass-transfer  episodes  occur.   In
particular,  for this  specific simulation  we obtain  7 mass-transfer
episodes.  However,  it is  important to realize  that this  number of
mass-transfer episodes is  actually a lower limit to  the real one, as
mass  transfer  is  determined  by  the numerical  resolution  of  the
simulations.   Figure \ref{orbits}  displays the  trajectories  of the
centers of  mass of both  white dwarfs during the  entire interaction.
As can  be seen in this  figure, both white  dwarfs describe initially
elliptical trajectories.  During the  first approach (at $t\simeq 400$
s) both white  dwarfs get very close each other,  but very little mass
is transferred from  the less massive white dwarf  to the most massive
one ---  see Fig.  \ref{deltam}.   As a consequence, the  less massive
white   dwarf,  although  tidally   distorted,  still   preserves  its
approximate   original   shape   and  describes   another   elliptical
orbit. This  first approach  is followed by  6 more close  passages in
which  both  the  maximum   and  the  minimum  distance  between  both
components  decrease.  Eventually,  during the  seventh  approach, the
less  massive white  dwarf dissolves  and  all its  remaining mass  is
removed.   As previously discussed  (see table  \ref{tab-hydro}), very
little mass  is ejected from  the system and, consequently,  the final
remnant contains most of the mass of the system.

Figure  \ref{deltam} shows  the mass  lost by  the less  massive white
dwarf, $-\Delta  M_2$, as a funtion  of time.  Note that  not all this
mass  is accreted  by  the  most massive  white  dwarf. As  previously
discussed, part  of this mass is  indeed accreted by  the most massive
one, part goes to form a debris region and a small fraction is ejected
from the system. It is  worth noting the first very weak mass-transfer
episode  at  $t\simeq  400$  s.   Note  as  well  that  in  subsequent
mass-transfer  episodes  the  less   massive  white  dwarf  losses  an
increasing fraction of  mass until it is totally  disrupted during the
seventh closest  approach.  Similarly, it is also  interesting to note
that  the  periods of  the  elliptical  orbits  of both  white  dwarfs
decrease for each one of  the succesive mass-transfer episodes --- see
also figure \ref{orbits}.  Figure \ref{deltam} also shows that no mass
is  transferred  between  two  succesive  close  approaches.   On  the
contrary,  the mass-transfer  rate is  non  zero only  during a  short
fraction  of time  during  the succesive  periastrons.   That is,  all
mass-transfer  episodes are  of very  short durations.   In  fact, the
detailed  SPH data  shows that  the less  massive white  dwarf  has an
ellipsoidal shape  until the last mass transfer  episode, during which
it  becomes totally  distorted.   This is  not  the case  of the  most
massive  white  dwarf, which  preserves  its  initial spherical  shape
during the entire interaction.

\begin{figure}
\begin{center}
\vspace{8cm}  
\includegraphics{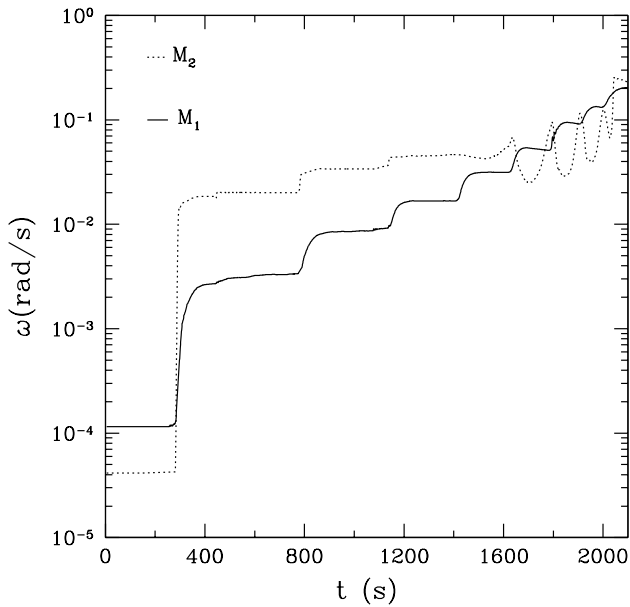}
\caption{Angular velocities  of the  interacting white dwarfs  for run
         number  4. The  angular velocity  of the  most  massive white
         dwarf  is shown  using a  solid line,  and that  of  the less
         massive one is depicted using a dotted line.}
\label{spins}
\end{center}
\end{figure}

Figure  \ref{spins} provides  additional details  of  the interaction.
Specifically, this figure shows that during the entire interaction the
white dwarfs  become partially synchronized. Obviously,  if both white
dwarfs were perfectly synchronized there would be a deformation of the
less  massive  one, and  this  deformation  would  be always  pointing
towards  the  line  connecting  the  centers of  mass  of  both  white
dwarfs. If  the rotation period of  the less massive  white dwarf were
smaller than the  orbital period of the system,  the tidal deformation
would be lagging behind the  line connecting both centers of mass, the
reverse  being also  true.   Consequently, there  would an  additional
gravitational  torque.  This torque  would have  an opposite  sense to
that  of   the  orbital  angular   momentum,  leading  to   a  perfect
synchronization of  the system. However, since  the synchronization of
the interacting  white dwarfs occurs  mainly during the  mass transfer
episodes it is clear that  synchronization is mainly the result of the
accretion stream connecting the less massive white dwarf with the most
massive   one.  Thus,   we  think   that   this  is   indeed  a   real
effect. However, given our  treatment of the artificial viscosity, see
\S  2, it is  possible as  well that,  although largely  reduced, some
shear viscosity  could be present during  the phases in  which no mass
transfer occurs  and, thus, some synchronization could  partially be a
spurious effect.

\subsection{Gravitational waveforms}

Since a sizeable amount of  mass is accelerated at considerable speeds
during the  interaction of the two  white dwarfs and  since the system
presents a  large degree of  asymmetry, we expect that  a considerable
amount of gravitational waves should be radiated. It is thus important
to characterize which would be  the gravitational wave emission of the
white dwarf interactions studied here and to assess the feasibility of
detecting them. 

To  compute  the  gravitational  wave  pattern,  we  proceeded  as  in
Lor\'en--Aguilar  et   al.   (2005).   In  particular,   we  used  the
weak-field quadrupole approximation (Misner et al. 1973). Higher order
terms of gravitational wave  emission could be included in calculating
the strains.  These terms  include the current-quadrupole and the mass
octupole.   It has been  shown (Schutz  \& Ricci  2001) that,  for the
first  of  these  to  be  relevant, an  oscillating  angular  momentum
distribution with a  dipole moment along the angular  momentum axis is
needed.   Consequently, in  our  calculations only  the mass  octupole
should  be  considered  in  the   best  of  the  cases.   Within  this
approximation, a term close to $v/c\sim 10^{-3}$ would be added to the
derived strains.   We have  found that for  the cases studied  here is
totally negligible, and thus we do not include it.

\begin{figure}
\vspace{13.5cm}  
\begin{center}
\includegraphics{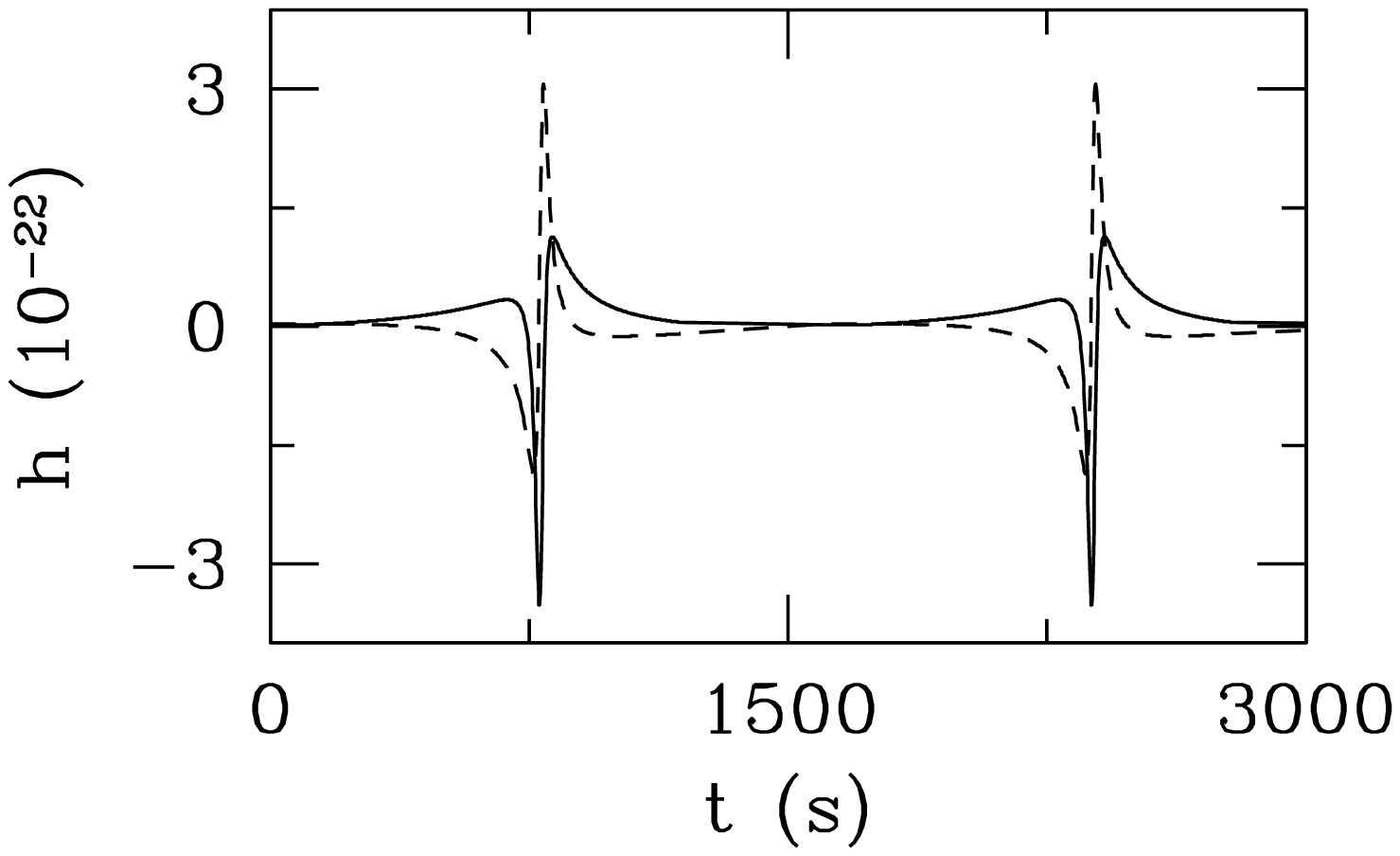}
\includegraphics{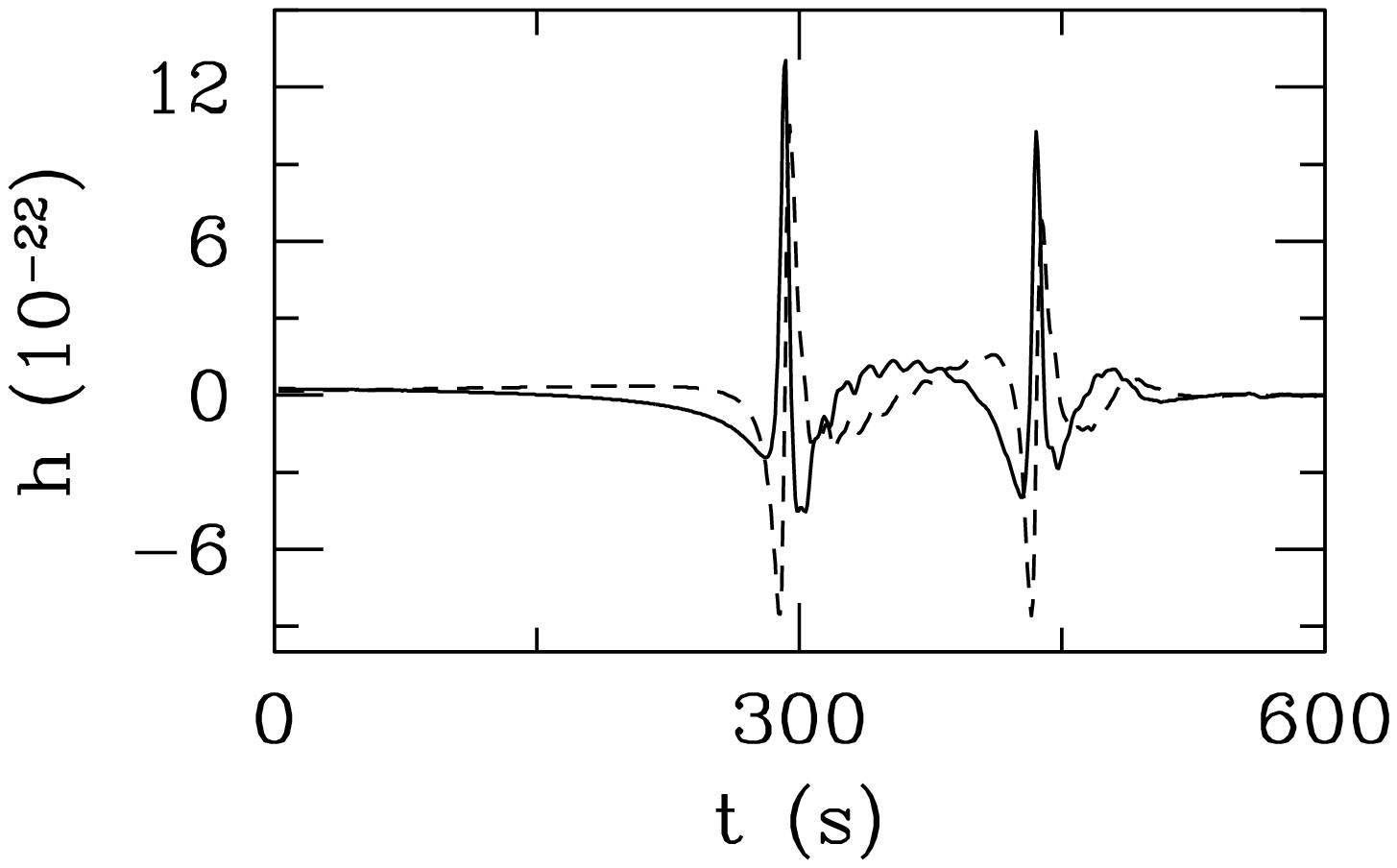}
\includegraphics{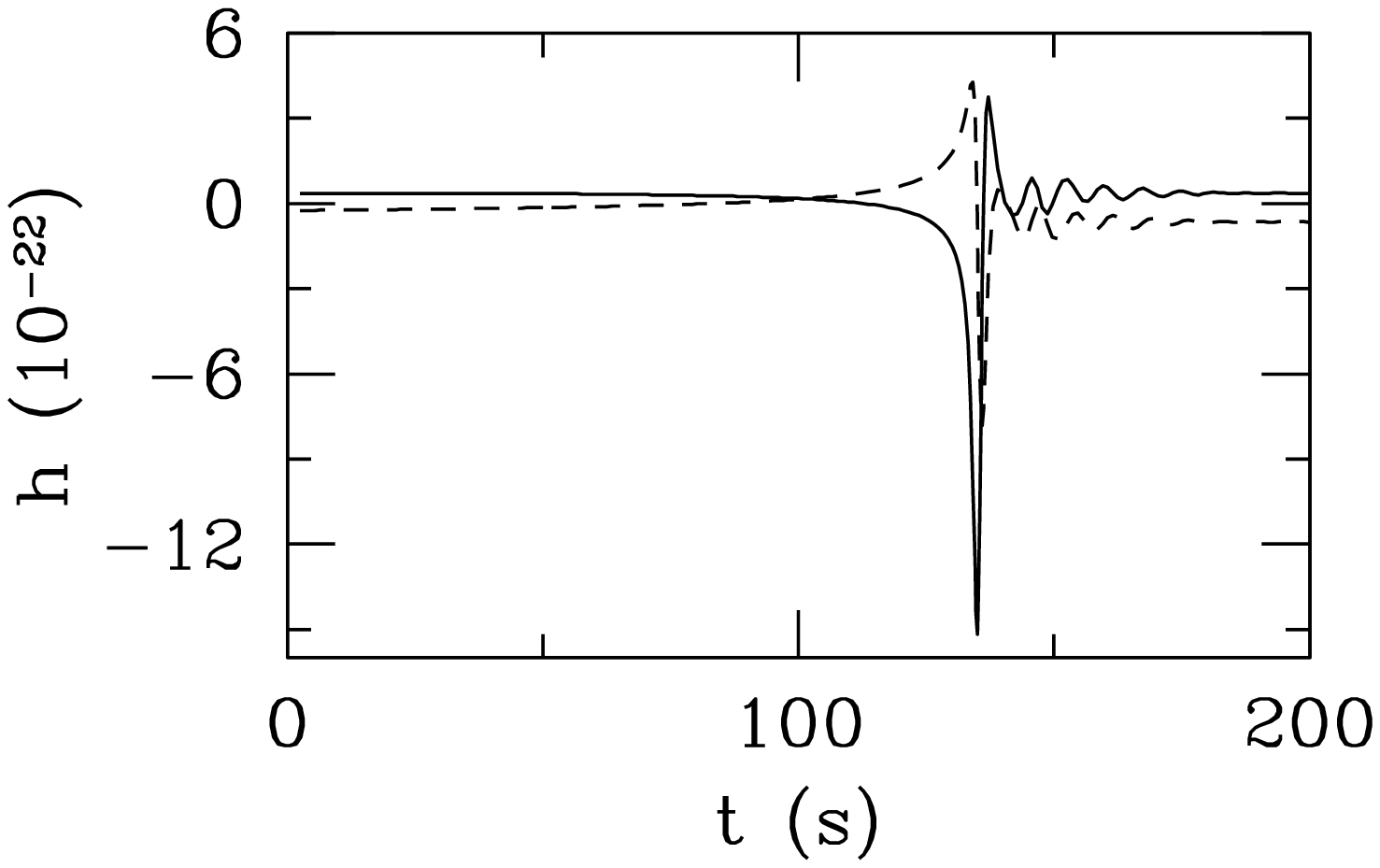}
\caption{Gravitational wave  emission from the collision of  a $0.6 \,
         M_{\sun}$ and  a $0.8 \,M_{\sun}$ white  dwarfs.  The initial
         distances  and velocities  are  from top  to bottom:  $y_{\rm
         ini}=0.6   R_{\sun}$  and   $v=150$~km/s,   $y_{\rm  ini}=0.3
         R_{\sun}$ and $v=175$~km/s and $y_{\rm ini}=0.1 R_{\sun}$ and
         $v=120$~km/s,  which lead  to the  survival of  the eccentric
         binary,   a  lateral  collision   and  a   direct  collision,
         respectively.   The   dimensionless  strains  $h_{+}$  (solid
         lines) and $h_{\times}$ (dashed  lines) are measured in units
         of  $10^{-22}$.   The source  is  located  at  a distance  of
         10~kpc.}
\label{fig:gwr}
\end{center}
\end{figure}

The three different behaviors  found in the previous sections directly
translate  in  the  form   of  the  dimensionless  strains  $h_+$  and
$h_\times$.  Examples of the gravitational waves released are shown in
figure~\ref{fig:gwr}. These gravitational  waveforms correspond to the
three cases  previously described  in \S 4.1  and depicted  in figures
\ref{orbital}, \ref{lateral} and  \ref{direct}.  Specifically, the top
panel  shows the  gravitational waveforms  in  the case  in which  the
eccentric binary white dwarf survives. As can be seen, the waveform is
perfectly periodic.  The middle  and bottom panels show, respectively,
the waveforms obtained for a lateral collision and a direct collision.
In  both   cases,  at  early   times  the  system  does   not  radiate
gravitational waves  because the  accelerations are very  small.  Once
the stars sufficiently approach  each other, the signal rapidly grows.
In  the  case  of a  lateral  collision  ---  middle panel  of  figure
\ref{fig:gwr} --- the gravitational  wave emission presents two peaks,
corresponding  to the  two phases  of rapid  mass  transfer previously
discussed.  Between these two peaks the structure of the gravitational
waveforms is rather complex. There are clearly several small peaks ---
a  behavior  which resembles  the  ring-down  phase  observed in  many
mergers of compact objects --- superimposed on a monotonous increasing
function. This occurs because in a lateral collision, before the final
merger,  the two  interacting white  dwarfs describe  a few  orbits of
decreasing separation, in which  some mass transfer occurs between the
two components  of the system. In  the case of a  direct collision ---
bottom panel  of figure \ref{fig:gwr}  --- the signal first  grows and
then  suddenly fades  away. Note  that  a ring-down  phase is  clearly
visible in  this case. It is important  to point out as  well that for
all the cases in which a  merger occurs the total energy radiated away
in  the form  of gravitational  waves  is similar,  regardless of  the
interaction strength --- see table \ref{tab-hydro}. In particular, for
run  3, which  has an  interaction  strength $\beta\sim  4$ the  total
gravitational  wave  energy  released  amounts  to  $E_{\rm  GW}\simeq
4.2\times  10^{40}$  erg,  whilst  for  run 12,  with  an  interaction
strength  5 times  larger  ($\beta\sim 20$),  the total  gravitational
energy radiated is $E_{\rm GW}\simeq 8.3\times 10^{40}$ erg.  Finally,
for the cases  in which the eccentric binary  system remains bound the
emission  of  gravitational waves  will  most  likely circularize  and
shrink their orbits.  Once this occurs, the gravitational wave strains
will present an almost periodical sinusoidal-like pattern.

Amongst  the  three types  of  gravitational  wave  signals, only  the
periodic ones  are, in  principle, good candidates  to be  detected by
LISA. However, the frequency of the gravitational emission obtained in
our  simulations  lies  far  from  the optimal  sensitivity  range  of
LISA. The orbital period of the binary systems is given by

\begin{align}
\tau^2 = \frac{\pi^2 \mu k^2}{2E^3}
\end{align}

\noindent  which  for  the cases  studied  here  is  of the  order  of
$5\times10^{-4}$~Hz.   This frequency  is  far from  the optimal  LISA
frequency band  wich lies around $10^{-2}$~Hz.   For the gravitational
wave emission of these systems to be detectable by LISA, we would need
to further  decrease the  total energy of  the system while  keeping a
high angular momentum in order for  the stars to do not collide.  That
is, the systems should circularize  their orbits.  It is clear as well
that  the   formation  of   systems  with  quasi-circular   orbits  by
gravitational capture  is rather unrealistic.   The only way  in which
this  could happen  is by  a  fine-tuned energy  and angular  momentum
transfer  in  the  gravitational  capture phase,  which  seems  rather
unlikely.   Thus, in  order for  this kind  of systems  to be  able to
radiate a  gravitational signal able  to be detected by  LISA, orbital
circularization  is  mandatory.   If  this circularization  is  mainly
driven by the emission of  gravitational waves, its time scale will be
given by  $\tau_{\rm GW}  = E/L_{\rm GW}$,  where $L_{\rm GW}$  is the
luminosity radiated as gravitational  waves, which is given by (Peters
\& Matthews 1963):

\begin{align}
L_{\rm GW}  &= \left[\frac{32}{5}\left(\frac{G^4}{c^5}\right)
M_1^2M_2^2(M_1+M_2)/a^5\right]f(\epsilon) \\
f(\epsilon)  &= \frac{1+\frac{73}{24}\epsilon^2+
\frac{37}{96}\epsilon^4}{(1-\epsilon^2)^{7/2}}
\end{align} 

\noindent Using  typical values we  obtain $\tau_{\rm GW}\sim  1$ Gyr,
which  is much smaller  than the  typical age  of a  globular cluster.
Thus,  we expect  that most  of  the systems  would have  circularized
orbits.  According  to the  recent work of  Ivanova et al.   (2006), a
typical cluster of $2\times 10^5\, M_{\sun}$ will have at least 3 LISA
binaries  at any  given moment  with periods  shorter than  a thousand
seconds. Thus, possibly, some of these binaries could be the result of
the  systems  studied here.   However,  a  direct  detection of  these
collisions is quite unlikely.


\subsection{Fallback luminosities}

Another  potential observational  signature of  the  mergers resulting
from the  collisions studied  here is the  emission from  the fallback
material in  the aftermath of some  of the close  encounters.  We have
already shown that  as a result of the collisions  of two white dwarfs
in some  of the cases  studied so far  a merger occurs.  We  have also
shown  in \S  4.2 that  the structure  of the  remnants consists  in a
central  compact object  surrounded by  either a  keplerian disk  or a
spherically symmetric cloud.  Most of the SPH particles of the disk or
cloud have circularized orbits.  However,  as it occurs in the mergers
of  double neutron  stars  or  white dwarfs,  some  material from  the
colliding white  dwarfs is found to  be in highly  eccentric orbits as
well.  After some  time, this material will most  likely interact with
the recently formed disk or cloud.  As discussed in Rosswog (2007) the
timescale for this is not  set by viscous dissipation but, instead, by
the  distribution  of eccentricities.   We  follow  closely the  model
proposed  by Rosswog (2007)  and we  compute the  accretion luminosity
obtained from the interaction of the material with high eccentricities
with  the  remnant  by  assuming  that the  kinetic  energy  of  these
particles is dissipated within the radius of the debris disk or cloud.
Our calculations  are done in  the reference system comoving  with the
central remnant.

\begin{figure}
\begin{center}
\vspace{8cm}  
\includegraphics{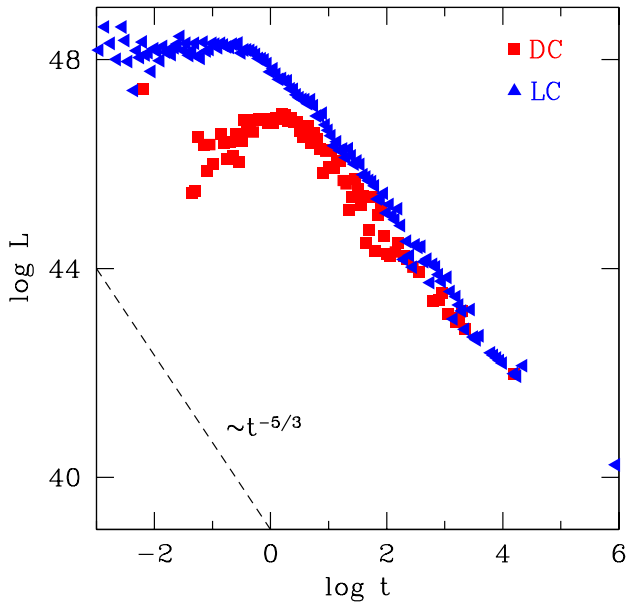}
\caption{Fallback accretion luminosity for  the cases of a direct (DC)
         and lateral  collision (LC).  The units of  time are seconds,
         whereas those of luminosities are erg/s. A straight line with
         slope 5/3 is shown for comparison.}
\label{lumin}
\end{center}
\end{figure}

In figure \ref{lumin} we plot the accretion luminosities as a function
of time for the fiducial  cases presented in figures \ref{lateral} and
\ref{direct}. We emphasize that  these luminosities have been computed
assuming  that the highly  eccentric particles  loose all  its kinetic
energy when interacting with the disk or cloud, for which we adopt the
radius obtained  by the  end of  our SPH simulations  and that  only a
fraction of  this energy will be  released in the  form of high-energy
photons. Thus, the  results shown in Fig. \ref{lumin}  can be regarded
as an  upper limit for  the actual luminosity of  high-energy photons.
Note that  although the luminosities are smaller  than those typically
obtained  for  the  merger  of  double neutron  stars  ---  which  are
typically of  the order  of $\sim 10^{52}$  erg/s --- they  are rather
high.  This is an important  result because it shows that observations
of high-energy  photons can help  in detecting the  gravitational wave
signal radiated by these systems.

It is important  to note as well that for  direct collisions the slope
of the  accretion luminosities differ  slightly from that  obtained in
the case of white dwarf  mergers ($\propto t^{-5/3}$).  The reason for
this is the distribution  of eccentricities of the accreted particles.
We start our discussion noting that in the analytical model of Rosswog
(2007) the central object  was assumed to be a  black hole, whereas we
obtain  a central  massive white  dwarf.   Moreover, in  the model  of
Rosswog (2007) there  are two key parameters, the  mass of the central
compact remnant  and the  geometry of the  surrounding disk  or cloud.
Simple numerical  experiments reveal that the precise  geometry of the
disk or cloud  only affects to the total  radiated energy, whereas the
mass  of  the the  central  object mainly  affects  the  slope of  the
fallback  luminosity.  These  dependances  are  also  present  in  the
calculations  of  Rosswog (2007),  see  his  figure  3. As  previously
mentioned, in our  calculations the central object is  a massive white
dwarf  surrounded by  a debris  region,  not a  point-like mass,  and,
accordingly,  we  compute  the  fallback  luminosity  using  the  real
gravitational  potential.   Additionally,  those particles  with  very
large eccentricities pass  very close to the central  white dwarf and,
thus, feel  a gravitational potential  of an extended source.   On the
contrary, and for the same  reasons, the gravitational well that those
particles  with  smaller  eccentricities  feel  resembles  that  of  a
point-like source.   Finally, the particles  with large eccentricities
reach their periastron at late times  because at the end of the merger
they  are located,  on  average, at  larger distances.   Consequently,
these particles  contribute to the  fallback luminosity at  late times
and  the contrary  holds  for particles  with smaller  eccentricities.
Thus,  at  late  times  we  expect  that the  slope  of  the  fallback
luminosity will differ from the  canonical value of $-5/3$, whilst for
early times we  expect that the this should be the  slope, and this is
indeed what we find.  As can  be seen in figure \ref{lumin}, for $\log
t <  3$ the  slope of  the fallback luminosity  is $\sim  -1.64$, very
close to  the canonical  value, while  for $\log t  > 3$  the fallback
luminosity can  be best approximated  by a $\sim t^{-1}$  curve. Note,
however,  that for the  case of  lateral collisions  the slope  of the
accretion luminosity presents the  classical $-5/3$ slope.  Indeed, in
this case the  resulting disk is very similar to  that obtained in the
merger of  white dwarfs, and  thus in this  case the results  are very
similar.

\section{Discussion and conclusions}

In this paper  we have studied the collisions  and close encounters of
two  white  dwarfs,  using  a  state  of  the  art  Smoothed  Particle
Hydrodynamics code.   Collisions between two  white dwarfs are  not as
frequent as  binary mergers.  However, as discussed  in Timmes (2009),
Rosswog  et al.  (2009)  and Raskin  et al.  (2009), they  most likely
occur in globular clusters and  the central regions of galaxies, where
the stellar  densities are very high.  The  collision time, $\tau_{\rm
coll}$,  adopting a Maxwellian  velocity distribution  with dispersion
$\sigma$, and  assuming a closest approach distance  $r_{\min} < 2R_*$
is (Binney \& Tremaine 1987):

\begin{equation}
\frac{1}{\tau_{\rm coll}} = 16\sqrt{\pi}n_{\rm WD}\sigma R_*^2
\left(1+\frac{v_{\rm esc}^2}{4\sigma^2}\right)
\end{equation}

\noindent  where  $n_{\rm WD}\simeq  10^4$  pc$^{-3}$  is the  typical
number density  of white  dwarfs in a  globular cluster,  $v_{\rm esc}
\simeq  4000$~km/s is  the  white dwarf  escape  velocity and  $\sigma
\simeq  5$~km/s is  the  relative velocity  dispersion  of both  white
dwarfs  in a  the globular  cluster,  which is  entirely dominated  by
gravitational  focusing. Consequently,  the rate  of collisions  for a
typical globular cluster is given by:

\begin{equation}
r_{\rm GC} \sim \frac{1}{2}\frac{n_{\rm WD}}{\tau_{\rm coll}} 
\; \frac{4}{3} \pi r_{\rm c}^3
\end{equation}

\noindent  where $r_{\rm  c}\sim 1.5$  pc is  the core  radius  of the
globular  cluster (Peterson  \& King  1975).  Adopting  the previously
mentioned typical values, we  obtain $r_{\rm GC}\sim 8\times 10^{-10}$
yr$^{-1}$.  Taking into account  that the density of globular clusters
is $n_{\rm GC}  = 4.2$ Mpc$^{-3}$ (Brodie \&  Strader 2006) as Rosswog
et al.  (2009)  did, we obtain an overall  rate of interactions $R\sim
3\times   10^{-10}$  Mpc$^{-3}$   yr$^{-1}$.   Thus,   although  these
interactions are not  very frequent, they are not  unlikely and, thus,
there is a possibility of detecting them.

Motivated by this we have  studied the collisions and close encounters
of  two otherwise  typical white  dwarfs  of masses  $0.8$ and  $0.6\,
M_{\sun}$, respectively,  for a broad range of  initial conditions and
employing a large number of SPH particles. Our initial conditions have
been chosen  in such a  way that a  close encounter or a  collision is
always guaranteed, and are summarized in Table \ref{outcome}.  We have
found  that the outcome  of the  interactions can  be either  a direct
collision,  a  lateral   collision,  in  which  several  mass-transfer
episodes may occur,  and finally the survival of  the eccentric binary
system  of two  white  dwarfs.   We have  characterized  the range  of
initial velocities  and distances --- or, alternatively,  the range of
energies  and angular  momenta ---  which lead  to each  one  of these
outcomes. We find that when  the distance between the two white dwarfs
at the closest  approach is smaller than $0.009\pm  0.002 \, R_{\sun}$
the  final outcome  is  a  direct collision,  when  it ranges  between
$0.009\pm  0.002\,  R_{\sun}$   and  $0.033\pm  0.004\,R_{\sun}$,  the
outcome  is a lateral  collision and  otherwise the  double degenerate
binary system survives.

In all the cases in which a collision is the result of the interaction
we  obtain that  little  mass  is ejected  during  the entire  merging
process and that  a central white dwarf surrounded  by a debris region
is  formed.  If  the collision  is  direct this  region has  spherical
symmetry,  whilst  if the  collision  is  lateral  we obtain  a  heavy
rotationally-supported  keplerian  disk.    In  both  cases  the  peak
temperatures  achieved  during   the  interaction  exceed  the  carbon
ignition temperature  and some nucleosynthesis  occurs. However, since
these high temperatures  are not achieved during long  periods of time
the   abundances  of   heavy  nuclei   are  not   large   (see  tables
\ref{tab-chem-disk} and \ref{tab-chem-corona}).  Naturally, the extent
to  which nuclear  burning proceeds  depends  on the  strength of  the
interaction, and  hence the  production of heavy  nuclei is  larger in
direct  collisions. Most of  the nuclear  reactions occur  when matter
from the  less massive white  dwarf is shocked  on the surface  of the
most massive one.  Consequently, we find that the maximum temperatures
of the  merged system occur  on a hot  corona around the  most massive
white dwarf.

We have  also paid special attention  to a specific case  of a lateral
collision in which several (up to 7) mass-transfer episodes occur.  We
have found  that mass-transfer only occurs during  the periastron, and
that at  each passage the  distance between the two  interacting white
dwarfs decreases, that  the mass lost by the  less massive white dwarf
increases,  and  that  there  is substantial  synchronization  of  the
system.

We  have  also computed  possible  observational  signatures of  these
events. Specifically, we have calculated the emission of gravitational
waves and the fallback luminosities in the aftermath of the merger. We
have  shown  that  it is  very  unlikely  that  LISA will  detect  the
gravitational waves radiated during  these interactions.  Only at very
late  phases,  when  the  orbits  are circularized,  the  emission  of
gravitational waves from those systems in which an eccentric binary is
initially formed is  possible. However, at these very  late stages all
the information about the close encounter is completely lost.  We have
also  computed  the  fallback  luminosities which  result  from  those
interactions in  which a merger  occurs.  We have found  that although
these  luminosities are somewhat  smaller than  those obtained  in the
merger of  a binary  white dwarf system  they are still  rather large,
allowing the future detection of these events.

Finally, we would like to emphasize  that our main aim was to study to
the post-capture scenario for a  fixed pair of masses of the colliding
white  dwarfs. Our  study differs  from  those of  Rosswog (2009)  and
Raskin  et al. (2009)  in the  adopted masses  of the  colliding white
dwarfs and  in the  initial conditions. This  is a consequence  of the
different  motivations of  all three  works.  Whereas  the  studies of
Rosswog (2009) and  Raskin et al. (2009) were  aimed at producing Type
Ia supernova  outbursts, our  work was focused  at studying  the tidal
disruption  of typical white  dwarfs in  globular clusters.   For this
reason both Rosswog (2009) and and Raskin et al. (2009) studied direct
collisions adopting  different masses  for the colliding  white dwarfs
than those adopted in this study.  Additionally, Rosswog et al. (2009)
placed the colliding white dwarfs  in a parabolic orbit and, moreover,
the total mass of the system was larger than Chandrasekhar's mass.  In
this sense,  all three studies are complementary  but, obviously, more
studies are needed to explore the full range of possibilities.

\section*{Acknowledgments}
Part   of   this   work    was   supported   by   the   MCINN   grants
AYA2008--04211--C02--01 and and AYA08--1839/ESP, by the European Union
FEDER funds and  by the AGAUR. We thank our  anonymous referee for his
very positive attitude, constructive criticism and useful suggestions.

\end{document}